\documentclass[aps,pra,a4paper,twocolumn,longbibliography,superscriptaddress]{revtex4-2}
\usepackage{amsmath}
\usepackage{amssymb}
\usepackage{mathtools}
\usepackage{graphicx}
\usepackage[space]{grffile}   % to include filenames with spaces
\usepackage{comment}
\usepackage{tikz}

\newcommand{\beq}{\begin{equation}}
\newcommand{\eeq}{\end{equation}}
\newcommand{\bei}{\begin{itemize}}
\newcommand{\eei}{\end{itemize}}
\newcommand{\ben}{\begin{enumerate}}
\newcommand{\een}{\end{enumerate}}

\newcommand{\be}{{\mathbf e}}

\usepackage{hyperref}
\hypersetup{
   pdfpagemode=None, 
   pdfstartpage=1,
   pdfmenubar=true,
   pdftoolbar=true,
   colorlinks = true,
   linkcolor=blue,
   citecolor=blue,
   urlcolor=blue,
   bookmarksopen=false
 }

\definecolor{darkblue}{rgb}{0.,0.24,0.51}
\definecolor{britishracinggreen}{rgb}{0.0, 0.26, 0.15}
\definecolor{darkgreen}{rgb}{0,0.60,.2}

\def\be{\begin{equation}}
\def\ee{\end{equation}}

\def\pp#1{\frac{\partial}{\partial#1}}
\def\pbyp#1#2{\frac{\partial#1}{\partial#2}}

\def\rf#1{(\ref{#1})}
\def\VEV#1{\left\langle\,#1\,\right\rangle}

\begin{document}
\title{Strongly interacting impurities in a dilute Bose condensate}
\author{Nikolay Yegovtsev}
\affiliation{Department of Physics and Center for Theory of Quantum Matter, University of Colorado, Boulder CO 80309, USA}
\author{Pietro Massignan}
\affiliation{Departament de F\'isica, Universitat Polit\`ecnica de Catalunya, Campus Nord B4-B5, E-08034 Barcelona, Spain}
\author{Victor Gurarie}
\affiliation{Department of Physics and Center for Theory of Quantum Matter, University of Colorado, Boulder CO 80309, USA}

	\date{\today}
	
\begin{abstract}
An impurity in a Bose gas is commonly referred to as Bose polaron. For a dilute  Bose gas its properties are expected to be universal, that is dependent only on a few parameters
characterizing the boson-impurity interactions. When boson-impurity interactions are weak, it has been known for some time that the properties of the polaron depend only on the scattering length
of these interactions. In this paper 
which accompanies and extends Ref. \cite{Yegovtsev2021} (where some of these results have already been reported)
we examine stronger boson-impurity interactions, keeping their range finite. We demonstrate that for attractive interactions between impurity and the bosons up to and including the unitary point of these interactions, all static properties of a Bose polaron in a dilute Bose gas can be calculated in terms of the scattering length and 
an additional parameter which characterizes the range of the impurity-boson interactions. We show that our approach to this problem is valid if this parameter does not deviate too much from the scattering length of intra-boson interactions, with the precise criterion given in the text. 
We produce explicit expressions for the energy and other properties of polaron for the case when the impurity-boson scattering length is tuned to unitarity, and we also provide the first correction away from it.

\end{abstract}
\maketitle

\section{Introduction}

The study of impurities in Bose and Fermi gases has a long history \cite{Landau1933,Froehlich1954,Feynman1955,Gross1962,Padmore1971,Astrakharchik2004}. An impurity, that is an atom distinguishable  from those forming the gas, binds with the atoms of the gas to form a quasiparticle often called a polaron. 
Polarons in ultracold Fermi gases, both weakly and strongly interacting, have been studied for quite some time \cite{Schirotzek2009,Chevy2010,Kohstall2012,Koschorreck2012,Massignan_Zaccanti_Bruun,Cetina2016,Scazza2017,Schmidt2018,Yan2018,Adlong2020}. Polarons in Bose gases 
have recently been getting significant attention
 \cite{Rath2013,Hu2016,Jorgensen2016,Yan2020}.

We would like to apply the expansion in powers of the gas parameter of the Bose gas to the problem of a dilute Bose gas interacting attractively and arbitrarily strongly with a single impurity.
We argue that this expansion is valid no matter how strong the boson-impurity interactions are, up to the unitary point, as long as the typical spatial extend  of impurity-boson interactions is not too short. 

The interactions among the bosons in the gas are taken to be repulsive, with the range roughly of the order of the scattering length of the intra-boson interactions, as always the case for
weak repulsive interactions. We further argue that for the purpose of calculations this range can be taken to zero without qualitative change in the properties of the polaron. 

This approach allows us to produce a method to calculate the energy of the Bose polaron at arbitrary negative scattering length characterizing impurity-boson interactions, including when it is infinity. 
At infinite and close to infinite scattering length, we derive explicit expressions for the energy and other properties of the polaron. 

To demonstrate the validity of our approach, we estimate the first subleading term in the expansion in powers of gas parameter. We show that this term is indeed small regardless of the strength of the boson-impurity interactions, under reasonable assumptions on their range. 
This offers strong support to the claim that the leading term correctly captures the behavior of the impurity when the gas is weakly interacting and the range of the impurity-boson interactions is not too short.

We consider a weakly interacting Bose gas with density $n$. Suppose a number of impurity atoms are introduced in this gas with density $n_I \ll n$. Let us first briefly consider the thermodynamics of this gas with impurities, following earlier 
work \cite{Massignan2005}. 

The free energy per unit volume of this gas will depend on both densities
$F(n, n_I)$. It is advantageous to introduce a ``mixed" thermodynamic potential
\be \label{eq:G}  G(\mu, n_I) = F - \mu n. \ee
Here $\mu$ is the chemical potential of the gas, $\mu = \partial F/\partial n$. To maintain constant the density of the gas far away from the impurities we work in the regime where $\mu$ is fixed and is independent of $n_I$. 

At small impurity density we expect that $G(\mu, n_I)$ will have a regular Taylor expansion in powers of $n_I$,
$$ G(\mu, n_I) \approx G(\mu, 0) + E(\mu) \, n_I + E_2(\mu) \, n_I^2  + \dots .
$$
Here $E(\mu)$ is clearly the energy cost of adding a single impurity, while $E_2(\mu)$ and higher order terms describe interactions among the impurities. 

Let us work at a very low density where interactions among the impurities can be neglected. The density of the gas kept at chemical potential $\mu$ with the impurity density $n_I$ can now be found according to
\be \label{eq:denc} n = - \pbyp{G}{\mu} = n_0 - n_I \pbyp{E}{\mu}.
\ee
where 
$$ n_0 = - \pbyp{G(\mu,0)}{\mu}
$$ is the density of the gas before impurities were introduced. 

Multiplying \rf{eq:denc} by the volume $V$ on both sides, and noting that $V n_I$ is the total number of introduced impurities, we find that the number of 
bosons trapped in the potential created by a single impurity is
\be \label{eq:trapped} N(\mu) = - \pbyp{E}{\mu}.
\ee
This is a powerful relationship which allows us to concentrate on calculating $E$, simultaneously also determining number of trapped bosons $N$.

The goal of this paper is to calculate $E(\mu)$ and $N(\mu)$. While these quantities can be expected to depend on the the details of the interactions among the bosons as well as boson-impurity interactions, and while from the point of view of matching the calculation with experiment it may be important to analyze this problem with the range of parameters relevant for experiment, 
we expect that the behavior of these functions will be universal to some degree in the limit where the strength of intra-boson interactions is taken to zero. 

We would like therefore to analyze the functions $E(\mu)$ and $N(\mu)$ in the limit of very weak interactions among bosons, while the interaction between the impurity and the bosons remain arbitrary. In particular, the interaction can be allowed to be taken to the unitarity limit where it is effectively the strongest. 

As could be expected, we find that in this regime the properties of polaron do not show any substantial dependence on the range of the boson-boson interactions. However, they do show substantial dependence on the 
range of the boson-impurity interactions. The results reported here 
are valid if the range $R$ of the boson-impurity interaction satisfies
\be \label{eq:conditions} ( n_0 a_B^3)^{1/4} \ll \frac{R}{a_B} \ll \frac{1}{\sqrt{n_0 a_B^3}}.
\ee
Here $a_B$ is the scattering length of the boson-boson interactions and $n_0$ is the density of the Bose gas. Weak interactions among bosons of course implies $n_0 a_B^3 \ll 1$. 
The first of these two inequalities are obtained by demanding that the gas remains weakly interacting everywhere including at the position of the impurity, while the second of these inequalities 
asks that the range of the potential remains smaller than the healing length of the gas. 

Note that
this excludes the direct comparison of our results to those obtained in the zero range regime where $R \rightarrow 0$ such as Ref.~\cite{Parish2021}. However, we expect that the conditions
\rf{eq:conditions} are satisfied in experiment,  since for realistic interactions $R/a_B \sim 1$.

One of our main results  is that in the regime of weak boson-boson interactions, with the range 
of boson-impurity interaction satisfying  the conditions \rf{eq:conditions}, 
$E(\mu)$ and $N(\mu)$ can be calculated as a function of the boson-impurity scattering length $a$. In Ref.~\cite{Yegovtsev2021} we showed that in the 
unitary limit where $a$ is taken to infinity, they take the analytic form asymptotically valid under the conditions \rf{eq:conditions}
\be  \label{eq:energyas}
E(\mu) = - \frac{3 (\pi n_0)^{2/3}}{2 m} \left( \frac{R}{a_B} \right)^{1/3},
\ee
\be  \label{eq:particleas2}  N(\mu) = \frac{R^{1/3}}{4 (\pi n a_B^4)^{1/3}}.
\ee
Here $m$ is the reduced mass of bosons and impurity defined below in Eq.~\rf{eq:reduced}. For the purpose of Eqs.~\rf{eq:energyas} and \rf{eq:particleas2}   $R$
needs to be  the quantitatively  defined.
Precise definition of $R$ is given below in Eq.~\rf{eq:Rdef}.

Both of these are asymptotically exact in the limit where $a_B$ is taken to zero. 

We also calculate $E$ and $N$ when $a$ deviates from the unitary limit. While in principle our methods allow us to find these quantities for arbitrary $a<0$, in practice expressions for those 
quickly become cumbersome, so we discuss only  the expansion of $E$ and $N$ in powers of $1/a$, given in Eq.~\rf{eq:parama}, as well as the behavior of these quantities for small $a$. 

The rest of this paper in organized as follows. In Section \ref{sec:two} we set up the problem and the techniques we will use to analyze it. Section \ref{sec:three} describes the solution
of the problem both at weak and strong (close to unitarity) impurity-boson attractive  interaction, with the simplifying assumption of the intra-boson interactions range  taken to zero. Section
\ref{sec:fluct} studies fluctuational corrections to the saddle point approximation employed in Section \ref{sec:three} and derives the conditions \rf{eq:conditions}. 
Section \ref{sec:four} discusses
the implication of finite range intra-boson interactions, with the conclusion that the finite range does not strongly affect the behavior of the polaron, unlike the range of the boson-impurity interaction. Section \ref{sec:flat} discusses  a particular technique, expansion about the unperturbed condensate, which, while appealing at a first glance, can be shown to fail as one approaches the unitary limit of impurity-boson interactions.  Section \ref{sec:seven} discusses Bose-impurity interactions which support a bound state, with the conclusion that the polaron behavior in this regime depends on the detailed functional form of these interactions. 
Section \ref{sec:conclusions} presents our conclusions. Finally, in the Appendix, we demonstrate that the technique of Section~\ref{sec:flat}, if its range of applicability is ignored, produces the expression \rf{eq:energyflat} which appeared before 
in the literature \cite{Shchadilova2016}, but which our analysis indicates is not applicable at strong boson-impurity interactions.

\section{Setting up the problem}
\label{sec:two} 
We begin with a number of bosons of mass $m_b$ which interact among themselves via a short-range weak interaction $V_{bb}$, as well as interacting with a single impurity of mass $M$ via
another potential $U$. The problem can be set up according to
$$ {H } = \sum_j \frac{{\bf p}_j^2}{2m_b} + \frac{{\bf P}^2}{2M} + \sum_{jk} V_{bb}({\bf x}_j-{\bf x}_k) + \sum_j U({\bf x}_j-{\bf X}).
$$
 Here ${\bf x}_j$ and ${\bf p}_j$ are the coordinates and the momenta of the bosons, while ${\bf X}$ and ${\bf P}$ are the coordinate and the momentum of the impurity. 
 
Before proceeding further, we would like to state explicitly  that everywhere in this paper in accordance with conventions common in quantum many-body literature we set $\hbar=1$, $k_B=1$. 

As was already exploited in the literature in this context \cite{Shashi2014,Drescher2020}, it is convenient to get rid of the impurity coordinate by performing the Lee-Low-Pines unitary transformation \cite{Pines1953} of the Hamiltonian. Define
$$ W = {\bf X} \cdot \sum_j {\bf p}_j.
$$ 
Straightforward algebra shows that
$$ e^{iW} H e^{-iW} = $$
$$
\sum_j \frac{{\bf p}_j^2}{2 m_b} + \frac{\left( {\bf p}_0 - \sum_j {\bf p}_j \right)^2}{2M} + \sum_{jk} V_{bb}({\bf x}_j-{\bf x}_k) + \sum_j U({\bf x}_j).  
$$
Here ${\bf p}_0$ is the conserved total momentum of the system. 
In this representation, the position of the impurity has effectively been set at the origin of the reference frame. 
 
We are now going to choose the interactions  among bosons to be zero ranged 
\be \label{eq:vbb} V_{bb}({\bf x}) = \lambda \, \delta({\bf x}). \ee This is a largely technical step which will simplify
further analysis of this problem. A very natural question then is whether a similar simplification can be used with the  boson-impurity interaction potential $U$. We will see later that shrinking the
range of $U$ to zero is a singular limit as the properties of the boson-impurity cloud crucially depend on the range of the potential $U$. At the same time, the dependence on the range of $V_{bb}$ is weak
and can be neglected. Below we will also explore how the finite range of $V_{bb}$ modify our conclusions to confirm that the dependence on this range is indeed weak. 
 
We are now in a good position to rewrite the Hamiltonian in the second quantized notation. It is natural to  choose $U$ to depend
on the distance $r$ to the impurity only, to find
\begin{eqnarray}  \label{eq:massi} {\cal H} & =& \int d^3 x \left( \frac{\nabla \bar \psi \nabla \psi}{2m} + \left( U(r) - \mu \right) \bar \psi \psi + \frac{\lambda}{2} \left( \bar \psi \psi \right)^2 \right) + \cr
&& \frac{\left( {\bf p}_0 +i \int d^3 x \, \bar \psi {\bf \nabla} \psi \right)^2}{2M}.
\end{eqnarray}
Here $\mu$ is the chemical potential of the gas, and $m$ is the reduced mass of the boson and impurity,
\be \label{eq:reduced} m = \frac{m_b M}{m_b+M}.
\ee
 The coupling constant $\lambda$ is related to the scattering length $a_B>0$ characterizing interactions among bosons by $$\lambda =  \frac{4 \pi a_B}{m}.$$ 
 
Throughout this paper  we concentrate on the very heavy impurity where $M$ is very large. In this limit, the term on the second line of Eq.~\rf{eq:massi}  can be entirely neglected. 
 We also note that in this limit $m_b=m$. We plan to discuss the effects of the finite impurity mass in a different publication. 
 
We arrive at a very concrete formulation of the problem we would like to solve. A single heavy impurity can be effectively represented  by a potential $U({r})$ it induces on the gas which can be thought of as centered in the origin of the reference frame. 
Thus Hamiltonian ${\cal H}$ of the gas with an infinitely heavy impurity is  simply given by Eq.~\rf{eq:massi} with $M$ taken to infinity, or  
\be \label{eq:ham} {\cal H} = \int d^3 x \left( \frac{\nabla \bar \psi \nabla \psi}{2m} + \left( U(r) - \mu \right) \bar \psi \psi + \frac{\lambda}{2} \left( \bar \psi \psi \right)^2 \right).
\ee

To study the problem given by Eq.~\rf{eq:ham} we rely on the  functional integration formalism. 
To set up the functional integral, we construct the coherent state imaginary time action
\be \label{eq:acc} S = \int_0^{1/T} d\tau  \, \left(  \int d^3 x \, \bar \psi \pbyp{\psi}{\tau} + {\cal H} \right),
\ee
where $T$ is temperature. We will eventually take it to zero in most of the calculations, but it is convenient to keep it finite in some of the intermediate steps. 
With the help of this action we write down the functional integral which allows us to calculate $G$ defined in Eq.~\rf{eq:G},
\be \label{eq:free} e^{- \frac{VG}{T}}   = \int {\cal D} \psi \, e^{-S}.
\ee
Here $V$ is the volume of the system.

The gas we consider here is weakly-interacting, which is well known to imply that
\be \label{eq:weakc}  n_0^{-1/3} \gg a_B. \ee Note that
  $n_0^{-1/3}$ is the mean interparticle spacing in the gas. 
The chemical potential $\mu$ can be used to define the healing length $\xi$ of the gas according to
\be \mu = 1/\left( 2 m \xi^2 \right).
\ee
In a weakly-interacting Bose gas $\mu = \lambda n_0$, thus the condition for weak interactions can also be written as \be \label{eq:weakcc} \xi \gg n_0^{-1/3}. \ee   
  
  %%%%%%%%%%%%%%%%%

  \section{Solution via Saddle Point Approximation} 
  \label{sec:three}

Starting with the functional integral defined by Eqs.~\rf{eq:ham}, \rf{eq:acc} and \rf{eq:free}, we apply the saddle point approximation to find the Gross-Pitaevskii (GP) equation describing
this Bose condensate, which reads
\be \label{eq:GP} 
- \frac{\Delta \psi} {2m} + U \psi + \lambda \left| \psi \right|^2 \psi= \mu \psi.
\ee
Given the solution of this equation $\psi$, the energy of the polaron can be deduced by the substitution of it into Eq.~\rf{eq:ham} and subtracting the energy of the condensate without impurity, to give
\be \label{eq:energy1} 
E = -\frac{\lambda}{2} \int d^3 x  \left( \left| \psi \right|^4 - n_0^2 \right).
\ee
At the same time, the number of particles trapped in the polaron can be found by evaluating
$$ N = \int d^3 x \left[ \left| \psi \right|^2 - n_0 \right].
$$
We note that if the potential does not vary much on the scale of $\xi$, then the GP equation can be solved using local density approximation, as is often done in the case where $U$ represents the smooth potential of a trap holding the condensate. However, we are interested in the opposite limit where the range of the potential is much smaller than $\xi$. 

It is natural to ask whether fluctuations about the Gross-Pitaevskii equation are needed to study the polaron. We will argue in Sec.~\ref{sec:fluct}
 that as long as the gas is weakly interacting
and the range of the impurity-boson potential satisfies the conditions \rf{eq:conditions}, 
the Gross-Pitaevskii equation gives a good approximation to the behavior of the polaron. 

Eq.~\rf{eq:GP} is nonlinear and at a first glance looks intractable. We now demonstrate that nevertheless its analytic solution is possible
as long as $R \ll \xi$.

\subsection{Analytic solution to the Gross-Pitaevskii equation in an external potential} 
We would first like to work with potential which is strictly zero beyond some length $r_c$, $U(r)=0$ for $r>r_c$ (we will later be able to also consider potentials extending all the way to infinity). 
We introduce 
$$ \phi = \frac{\psi}{\sqrt{n_0}}.$$ In terms of this dimensionless condensate density function, Gross-Pitaevskii equation becomes
\be \label{eq:GP1} - \frac{\Delta \phi}{2m} + U \phi = \mu \phi \left( 1 - \left| \phi \right|^2 \right).
\ee
Since we are looking for the lowest energy solution, those will be real valued and spherically symmetric. 

We analyze the Eq~\rf{eq:GP1} by introducing a small parameter \be \epsilon = \frac{r_c}{\xi} \ee and constructing its solution as an expansion in powers of this parameter.
As a first step, it is convenient to split the range of $r$ into $0 \le r \le r_c$ and $r_c \le r < \infty$. In the first interval we introduce 
$$ y = \frac{r}{r_c}, \ \phi = \frac{\chi(y)}{y}.
$$ 
$\chi(y)$ satisfies
\be \label{eq:GPrad1} 
-\frac{d^2 \chi}{dy^2}  + 2m r_c^2 U \chi = \epsilon^2 \left( \chi - \frac{\chi^3}{y^2} \right)
\ee
on the interval $y \in [0, 1]$,
as well as $\chi(0)=0$. 
In the second interval we introduce  
$$ z=\frac r {\xi}, \ \phi = 1+\frac {u(z)}{z},
$$
to find
\be \label{eq:GPrad2} \frac{d^2 u}{dz^2} -2u = 3 \frac{u^2}{z} + \frac{u^3}{z^2}
\ee
on the interval $z \in [\epsilon, \infty]$, 
where $u \rightarrow 0$ when $z \rightarrow \infty$. 
We need to solve Eqs.~\rf{eq:GPrad1} and \rf{eq:GPrad2}, matching the behavior of their solutions at the boundary $r=r_c$. 

\subsubsection{Weak potential}
Let us first examine the case of weak attractive potential with a small scattering length $a<0$. We solve Eq.~\rf{eq:GPrad1} in the interval $0 \le y \le 1$ neglecting its right hand side as it contains a small parameter $\epsilon$. Then Eq.~\rf{eq:GPrad1} reduces to a Schr\"odinger equation in the potential $U$ at zero energy, 
\be \label{eq:sch}  -\frac{d^2 \chi}{dy^2}  + 2m r_c^2 U \chi =0, \ee
whose solution $\chi_0$ must satisfy $\chi_0(0)=0$. At $y>1$ the potential $U$ is zero, so $\chi$ must be a linear function.
Bethe-Peierls boundary conditions fix the form of this function to be
\be \label{eq:BP}  \chi_0(y) =  \frac{\alpha}{1- \frac{r_c}{a} } \left( 1- \frac{r_c}{a} y \right), \ y>1,
\ee 
where $a$ is the scattering length in the potential $U$ and $\alpha$ is the normalization of the solution chosen in such a way  that \be \chi_0(1)=\alpha. \ee

We can rewrite Bethe-Peierls boundary conditions in a convenient way by observing that they are equivalent to
\be \label{eq:BPd}  \chi'_0(1)  = \frac{\alpha}{1-\frac a {r_c}}.
\ee
We could continue to solve Eq.~\rf{eq:GPrad1} by means of successive approximations. The next correction $\chi_1$ to the solution satisfies
\be \label{eq:secondapp}
-\frac{d^2 \chi_1}{dy^2}  + 2m r_c^2 U \chi_1 = \epsilon^2 \left( \chi_0 - \frac{\chi^3_0}{y^2} \right).
\ee
$\chi_1$ is a small correction to $\chi_0$ since $\epsilon \ll 1$. We will see that $\chi_1$ becomes important close to the unitary point where $a \rightarrow \infty$. Then $\chi'_0(1) = 0$ and 
it is $\chi_1$ which produces a nonzero contribution to the derivative. But right now we are interested in small $a$. We will see a little later that we do not need to construct $\chi_1$ 
in the lowest approximation in $\epsilon$ in that case. 

Now we solve Eq.~\rf{eq:GPrad2} neglecting its right hand side  to find
\be \label{eq:la} u = A e^{-\sqrt{2} z}. \ee 
We can construct corrections to it by means of successive approximations. Those will be small as long as $A$ is small, which as we will see in a second is the case here. So for now we neglect those
corrections.   

Matching amplitudes and derivatives of $\chi(y)$ and $u(z)$ at $z=\epsilon$, or correspondingly $y=1$, produces
\be \label{eq:weakmatch} 
A = \epsilon(\alpha-1), \ -\sqrt{2} A = \frac {\alpha} {1-\frac a {r_c}}   - 1. \ee
Taking into account that $\epsilon \ll 1$, we find
\be \label{eq:solw}  A = -\frac{a} {\xi}, \ \alpha = 1-\frac {a}{r_c}.
\ee 

Let us now examine if the terms neglected to arrive at this solution are indeed small. We solve \rf{eq:GPrad1} by successive approximations, plugging $\chi_0$ into the right hand side of Eq.~\rf{eq:GPrad1} and producing a correction
$\chi_1$. If $\left| a \right| < r_0$ then both $\chi_0(1)$ and $\chi'_0(1)$ are of the order of $1$ while $\chi_1$ will be of the order of $\epsilon^2 \ll 1$ and can be neglected. It gets more interesting if $\left| a \right| >r_c$. Then 
$\chi_0(1) =\alpha \sim a/r_0$, while $\chi'_0(1) \sim 1$. At the same time, $\chi_1 \sim \epsilon^2 (a/r_c)^3$. The magnitude of this had better be smaller than $1$, so that the contribution of $\chi'_1(1)$ to the derivative of $\chi$ could be neglected. This condition gives
$\epsilon^2 {\left| a\right|^3}/{r_c^3} \ll 1$, or equivalently
\be \label{eq:weaks}  \left| a \right|^3 \ll  \xi^2 r_c.
\ee
This is the condition to neglect the right hand side of Eq.~\rf{eq:GPrad1}. We will see later that this condition signifies the transition from weak impurity potential, where Eq.~\rf{eq:weaks}  holds true, to the
strong potential including the unitary point $a \rightarrow \infty$, where it is violated. 

Under the condition \rf{eq:weaks} $A \ll 1$, so obviously we can indeed neglect the right hand side of \rf{eq:GPrad2} as we did above.

We can now plug our solution into Eq,~\rf{eq:energy1}. 
The integral in this formula needs to be split into two parts, $0 < r< r_c$ and $r_c < r$. It is easy to see that the contribution of the interval $0<r<r_c$, under the condition \rf{eq:weaks}, is negligible,
so we only need to integrate over $r>r_c$ with $u=A \exp(-\sqrt{2} z)$. 
Performing the integration  and again taking into account Eq.~\rf{eq:weaks} to get rid of some terms suppressed relative to the main contribution we find
\be \label{eq:energyweak} E= \frac{2 \pi n_0 a}{m}.
\ee
This is the well known result for the energy of the polaron at weak scattering length $|a| \ll \xi$. We now see that the validity condition for this to hold true is Eq.~\rf{eq:weaks}.

\subsubsection{Strong potential}
Suppose now that the potential $U$ is made more attractive so that its scattering length increases, violating the condition \rf{eq:weaks} and eventually reaching infinity at the unitary point. We can follow the same strategy to obtain the solution in this case. The new element  is that Bethe-Peierls boundary conditions now imply $\chi'_0(1)=0$, so we need to solve Eq.~\rf{eq:GPrad1} perturbatively, using its right hand side as a perturbation, to find nonzero $\chi'_1(1)$ contributing to $\chi'(1)$. Same goes for Eq.~\rf{eq:GPrad2}. 

In case when $a$ was small we found earlier that the amplitude $\chi_0(1) = \alpha$, $\alpha = 1- a/r_c$.  This is of the order of $1$ when $a$ is very small, but starts growing as $|a|$ is increased. 
When $|a|^3 \sim \xi^2 r_c$, $\alpha \sim \epsilon^{-2/3}$. We will see that $\chi(1)$ remains of the order of $\epsilon^{-2/3}$ even as $a$ is taken all the way to infinity, so it is convenient
to introduce the notation
$$ \beta= \alpha \epsilon^{2/3}, $$ 
so that
\be \label{eq:within} \chi_0(y) = \frac{\beta} {\epsilon^{2/3}} v(y). \end{equation} Here $v$ is the solution of the Schr\"odinger equation
\be \label{eq:sch25}  -\frac{d^2 v}{dy^2}  + 2m r_c^2 U v =0
\ee
normalized so that $v(1)=1$. $v'(1)=0$ since $U$ is tuned to unitarity. $\beta$ is thus a yet unknown $\epsilon$-independent normalization coefficient.

We will need a correction to this which satisfies
\be 
\label{eq:p1m} -\frac{d^2 \chi_1}{dy^2}  + 2m r_c^2 U \chi_1 = -\epsilon^2  \frac{\chi_0^3}{y^2}.
\ee
The term $\epsilon^2 \chi_0$ from the right hand side of Eq.~\rf{eq:secondapp} goes as $\epsilon^{4/3}$ and can be neglected. At the same time, we see that $\chi_1$ is
of the order of $\epsilon^0$. Solving  Eq.~\rf{eq:p1m} gives
$$ \chi_1 = \beta^3 v(y) \int_0^y \frac{ds}{v^2(s)} \int_0^{s} \frac{dt \, v^4(t)}{{t}^2}.
$$
Putting it together produces
\be \label{eq:seconds} \chi = \frac{\beta}{\epsilon^{2/3}}  v(y) + \beta^3 v(y) \int_0^y \frac{ds}{v^2(s)} \int_0^{s} \frac{dt \, v^4(t)}{{t}^2} + \dots.
\ee
The next term $\chi_2$ which can be obtained by continuing successive approximations goes as $\epsilon^{2/3}$. We will not need it here, but note that it will
have an even more complicated dependence on $v$ and by extension on features of the potential $U(r)$ than the already obtained term $\chi_1$.

From this solution we find that
\be \label{eq:bb1m} \chi(1) = \frac{\beta}{\epsilon^{2/3}} + \mathcal{O}(1),  \quad \chi'(1) = \beta^3 c + \mathcal{O}(\epsilon^{2/3}),
\ee where the dimensionless coefficient $c$ is defined via
\be
 \quad c= \int_0^1 \frac{dy \,
v^4(y)}{y^2}.
\ee

Now we turn our attention to Eq.~\rf{eq:GPrad2}. Its solution $u(z)$ needs to be matched with the boundary conditions \rf{eq:bb1m}. Easy to verify that 
these boundary conditions imply
\be \label{eq:bb2m} u(\epsilon) = \beta \epsilon^{1/3} + \mathcal{O}(\epsilon), \quad u'(\epsilon) = - 1 + \beta^3 c + \mathcal{O}(\epsilon^{2/3}).
\ee

Eq.~\rf{eq:GPrad2} differs from Eq.~\rf{eq:GPrad1} in that its nonlinear terms do not have an explicit factor of $\epsilon$ in front of them. We will nonetheless solve 
Eq.~\rf{eq:GPrad2} by means of successive approximations, and verify later that this is a legitimate approach. Without its right hand side, the solution to Eq.~\rf{eq:GPrad2} reads
as before,
\be \label{eq:uzerom} u_0 (z) = A \, e^{- \sqrt{2} z}.
\ee We can plug it into the right hand side of Eq.~\rf{eq:GPrad2}, however we will follow a slightly different procedure. 
We use Eq.~\rf{eq:uzerom} to rewrite Eq.~\rf{eq:GPrad2} as an integral equation via a standard procedure. This involves solving the auxiliary equation
$$ \frac{d^2 u}{dz^2} -2u = g(z)
$$
with arbitrary given $g(z)$, then substituting the actual right hand side of Eq.~\rf{eq:GPrad2}. We find
\begin{widetext}
\be  \label{eq:itm}  u(z) = A \, e^{-\sqrt{2} z} +  \frac{e^{-\sqrt{2} z}}{2 \sqrt{2}} \int_z^\infty \frac{ds \, e^{\sqrt{2} s}}{s} \left(3 u^2 (1+u') + \sqrt{2} u^3 \right) -  \frac{e^{\sqrt{2} z}}{2 \sqrt{2}} \int_z^\infty \frac{ds \, e^{-\sqrt{2} s}}{s} \left(3 u^2 (1+u') - \sqrt{2} u^3 \right). 
\ee
\end{widetext}
We now use this equation to calculate $u(\epsilon)$ and $u'(\epsilon)$ in perturbative expansion in powers of $\epsilon$. Anticipating that the leading
behavior $A \sim \epsilon^{1/3}$, as should be clear from comparing Eq.~\rf{eq:uzerom} and Eq.~\rf{eq:bb2m}, we iterate Eq.~\rf{eq:itm} by plugging $u_0(z)$ into the right hand side of Eq.~\rf{eq:itm}. The resulting integrals can be computed in terms of Gamma functions and expanded in powers of $\epsilon$.  We omit rather lengthy algebra involved, and just state that this allows us to evaluate $u(\epsilon)$ to find
\be \label{eq:vam} u(\epsilon) = A + \frac{3 \ln 3 }{2 \sqrt{2}} A^2 - A^3 \ln \epsilon + \mathcal{O}(\epsilon).
\ee
We also evaluate $u'(\epsilon)$. Differentiating Eq.~\rf{eq:itm} gives
\begin{widetext}
\be \label{eq:derrm} u'(z) = -\sqrt{2} A e^{-\sqrt{2} z} - \frac{u^3}{z}- \frac{e^{-\sqrt{2} z}}{2} \int_z^\infty \frac{ds \, e^{\sqrt{2} s}}{s} \left(3 u^2 (1+u') + \sqrt{2} u^3 \right) -  \frac{e^{\sqrt{2} z}}{2 } \int_z^\infty \frac{ds \, e^{-\sqrt{2} s}}{s} \left(3 u^2 (1+u') - \sqrt{2} u^3 \right).
\ee
\end{widetext}
We can again substitute $u_0(z)$ into the integrals on the right hand side of Eq.~\rf{eq:derrm}, to find
\be \label{eq:derr2m} u'(\epsilon) = -\sqrt{2} A - \beta^3 + 3 A^2 \ln \epsilon + \mathcal{O}(\epsilon^{2/3}).
\ee 
Here we took advantage of the boundary conditions \rf{eq:bb2m} which tell us that $u(\epsilon) = \beta \epsilon^{1/3}$ within the accuracy that we work with. 

Combining Eq.~\rf{eq:vam} and Eq.~\rf{eq:derr2m} with Eq.~\rf{eq:bb2m} gives 
\be \label{eq:syssm}  \begin{matrix} A + \frac{3 \ln 3 }{2 \sqrt{2}} A^2 - A^3 \ln \epsilon & = & \beta \epsilon^{1/3}, \cr
-\sqrt{2} A - \beta^3 + 3 A^2 \ln \epsilon & = &  - 1 + \beta^3 c. \end{matrix}
\ee
We now need to solve these equations for $A$ and $\beta$ perturbatively, in powers of $\epsilon$. Introduce 
$$ \delta = \frac{\epsilon}{1+c}.
$$ The solution to Eq.~\rf{eq:syssm} reads
\begin{eqnarray} \label{eq:param}
A & = & \delta^{1/3} - \left(  \frac{3 \ln 3}{2 \sqrt{2}} + \frac{\sqrt{2}}{3} \right) \delta^{2/3} + 2\delta \ln \delta + \mathcal{O} ( \delta), \cr
 \beta \epsilon^{1/3}  &=& \delta^{1/3} - \frac{\sqrt{2}}{3} \delta^{2/3} + \delta \ln \delta + \mathcal{O}(\delta).
\end{eqnarray}
We can now use the parameters we obtained in this way to calculate the energy and the particle number of the polaron. It turns out to be 
technically easier to calculate the
particle number first and then use Eq.~\rf{eq:trapped} to find the energy, which is the strategy we will follow here.

The excess number of particles due to the polaron is given by
$$ N = \int d^3 x  \left[ \left| \psi \right|^2 - n_0 \right] = 4 \pi n_0 \xi^3 \int_0^\infty z^2 dz \left[ \phi^2-1 \right].
$$
It is natural to split the integral over $z$ into two intervals, from $0$ to $\epsilon$ and from $\epsilon$ to infinity. Now the contribution of the first  interval can
be safely neglected. Indeed, it gives
$$ \int_0^{\epsilon} z^2 dz \left[ \phi^2-1 \right] = \epsilon^3 \int_0^1 dy \left( \frac{\chi^2}{y^2}-1 \right) \sim \epsilon^{5/3}.
$$
Here we used that $\chi(y) \sim 1/\epsilon^{2/3}$. This contribution is very small and exceeds the accuracy in $\epsilon$ with which we were doing our calculations. This also indicates that the bulk of the particles bound by the impurity are located farther than distance $r_c$ away from the impurity. 

The contribution of the second interval gives
\be \label{eq:pp3m} \int_\epsilon^{\infty} z^2 dz \left( \left( 1+ \frac{u}{z} \right)^2 - 1\right).
\ee
To evaluate this integral we again iterate  Eq.~\rf{eq:itm} once, to find $u$ up to the terms of the order of $A^2$, and substitute that into Eq.~\rf{eq:pp3m}. The result of this
evaluation is
\be \label{eq:partFinals} N = 4 \pi n_0 \xi^3 \left(  \delta^{1/3} - \frac{5}{3 \sqrt{2}} \delta^{2/3} + 2 \delta \ln \delta  + \dots \right).
\ee
Thus we evaluated the number of particles trapped in the polaron up to terms of the order of $\delta \ln \delta$. To go beyond this order, starting from
terms of the order of $\delta$ and beyond represented by the dots above, we would need to go beyond the terms presented in Eq.~\rf{eq:seconds}. We expect that this will produce terms which depend on the features of the potential other than the coefficient $c$.

To construct the energy of the polaron, it is easiest at this stage to take advantage of Eq.~\rf{eq:trapped}. The subtlety in evaluating the derivative there is that the particle number $n_0$ as well as $\xi$ have to be traded for $\mu$ before differentiating.    Doing the algebra we arrive at 
\be\label{eq:energyFinals}
E = - \frac{ \pi n_0 \xi}{m} \left( 3 \delta^{\frac 1 3} - 2 \sqrt{2} \delta^{\frac 2 3} + 4 \delta \ln \delta + \dots \right).
\ee
This constitutes the answer that we seek, the energy of the polaron at unitarity as an expansion in powers of $\delta \sim r_c/\xi$.

Finally, let us examine $\delta= \epsilon/(1+c)$ in a little more detail. From the definition of $c$ given in Eq.~\rf{eq:bb1m} we can write 
$$ 1+ c = 1+ r_c \int_0^{r_0} \frac{dr \, v^4}{r^2} =r_c  \left( \int_{r_c}^\infty \frac{dr} {r^2} + \int_{0}^{r_c} \frac{dr \, v^4}{r^2} \right).
$$
 $v$ is the solution of the Schr\"odinger equation with the potential tuned to unitarity, so that $v'(r_c)=0$. Since it is normalized such that $v(r_c)=1$, it will naturally satisfy $v(r)=1$ for all $r\ge r_c$. 
Therefore we can rewrite this as
$$ 1+c = r_c \int_0^\infty \frac{dr \, v^4}{r^2}.
$$
Now
$$ \delta= \frac{\epsilon}{1+c} = \frac{1}{\xi  \int_0^\infty \frac{dr \, v^4}{r^2} }.
$$
It is now convenient to define
\be \label{eq:Rdef} R^{-1} =  \int_0^\infty \frac{dr \, v^4}{r^2}  = \int \frac{d^3 x} { 4 \pi}  \,  \left| \psi_0 \right|^4, 
\ee
where $\psi_0 = v/r$ is the solution of the Schr\"odinger equation
$$ - \frac{\Delta}{2m} \psi_0 + U \psi_0 =0.
$$
$R$ constitutes a properly defined range of the potential, finite even for potentials which extend all the way to infinity, which correctly captures its extent for the purposes of solving the polaron problem.

 This gives us a definition
\be \label{eq:deltadef}  \delta = \frac{R}{\xi}.
\ee
At this stage $r_c$ drops from the equations and no longer needs to be finite. It
 can be taken to
infinity if desired, with the answer for the energy of the polaron \rf{eq:energyFinals} as well as the number of particles trapped in the polaron \rf{eq:partFinals} expressed entirely in terms of $\delta=R/\xi$.

\subsubsection{Perturbing away from unitary point}
\label{sec:largeacor}
We can go further and generalize the above analysis to account for the small deviations away from unitarity using a $1/a$ expansion. 
To accomplish this, it is convenient to parametrize $a$ by a new variable $\eta$ according to
\be \eta = \frac{ r_c^{1/3}\xi^{2/3}}{a}. \ee
Declaring $\eta \ll 1$ is equivalent to saying that we are in the strongly interacting regime (compare with the condition \rf{eq:weaks} which is violated once we cross over from weak to strong potentials
as explained earlier). 
We then proceed to work out the expansion in powers of $\eta$. The advantage of this reparametrization is due to form that Bethe-Peierls boundary conditions take. Indeed, Eq.~\rf{eq:BPd} at
large $a$ where
close to unitarity we should write $\alpha = \beta/\epsilon^{2/3}$ imply
\be \frac{\beta}{\epsilon^{2/3}(1-\frac{a}{r_c})} \approx -\frac{\beta r_c}{\epsilon^{2/3} a} =- \beta \eta.\ee

Because $r_c/a \sim \epsilon^{2/3}$ the correction to the nonlinear term and  to the  RHS of the top equation in (\ref{eq:syssm}) that arise from correction to $v(y)$ in Eq.~(\ref{eq:seconds})  are of the higher order in power of $\epsilon$ and thus can be neglected. Thus slighly away from unitarity the only modification to the formalism presented in the previous section is the addition of the $-\beta\eta$ term to RHS of the second equation in (\ref{eq:syssm}):
\be \label{eq:syssma}  \begin{matrix} A + \frac{3 \ln 3 }{2 \sqrt{2}} A^2 - A^3 \ln \epsilon & = & \beta \epsilon^{1/3}, \cr
-\sqrt{2} A - \beta^3 + 3 A^2 \ln \epsilon & = &  - 1 - \beta \eta + \beta^3 c. \end{matrix}
\ee

Notice that now in order to find the leading term in the expansion of $\beta$ in powers of $\epsilon$, we have to solve the cubic equation:
$$ \beta_0^3(1+c)-\beta_0\eta -1 =0.
$$
The number of real roots of this equation depends on the value of the discriminant $\Delta = 4(\frac{\eta}{1+c})^3-\frac{27}{(1+c)^2}$. If $\Delta >0$, there are three real roots, and if $\Delta <0$, only a single real root. The critical value of $\eta$ is obtained by setting $\Delta = 0$ and gives $\eta_c = (\frac{27(1+c)}{4})^{1/3}>0$. A positive value of $\eta$ corresponds to a positive value of the scattering length, which means that we are in the regime where the potential has a single bound state. It is known that when a potential admits $\nu$ number of bounds states, then the Gross-Pitaevskii equation can have up to $2\nu +1$ solutions \cite{Massignan2005}. This means that the approach outlined above can in principle be used to construct other solutions for the Gross-Pitaevskii equation, but since we are interested only in the ground state physics, we do not consider them here. Instead, just as already elaborated above we are going to assume that $\left| \eta \right| \ll 1$, so that we are in the regime where there is only a single solution exists and we can do expansion in $\eta$ on top of the expansion in $\epsilon$. The first order correction in  $1/a$ corresponds to the first order correction in $\eta$. Note that the case of large and positive scattering length corresponds to the presence of the bound state with energy $E_\text{binding} = -\frac{1}{2ma^2}$. 
Since  $r_c/a = \eta \epsilon^{2/3}$, in order to capture $1/a^2$ effects we would need to construct the further corrections to the matching equations that would include higher powers of
$\epsilon$ and therefore depend on other details of the potential. Therefore, in the lowest order approximation in $\eta$ that we are going to employ, we are not sensitive to the presence of the bound state
even if $a$ is positive. 

Section \ref{sec:seven} further elaborates on what we expect to happen in the regime where $a>0$ and a bound state is present once we move away from the unitary point. In particular, we expect that in this regime the behavior of the 
polaron strongly depends on the details of the impurity-boson potential $U$.

Let us go back to staying in the vicinity of the unitary point by demanding $| \eta | \ll 1$. Since $\frac{\eta}{(1+c)^{1/3}} = \frac{\delta^{1/3}\xi}{a}$, solution to the system (\ref{eq:syssma}) reads:

\begin{widetext}
\begin{eqnarray} \label{eq:parama1}
A & = & \delta^{1/3}\left(1+ \frac{\delta^{1/3}\xi}{3a} \right) - \left(  \frac{3 \ln 3}{2 \sqrt{2}} + \frac{\sqrt{2}}{3} \right) \delta^{2/3} - \frac{\ln 3}{\sqrt{2}} \cdot \frac{\xi \delta}{a} + 2\left(1+ \frac{2\delta^{1/3}\xi}{3a} \right) \delta \ln \delta  + \mathcal{O} ( \delta), \cr
 \beta \epsilon^{1/3}  &=& \delta^{1/3}\left(1+ \frac{\delta^{1/3}\xi}{3a} \right) - \frac{\sqrt{2}}{3} \delta^{2/3} + \left(1+ \frac{\delta^{1/3}\xi}{3a} \right) \delta \ln \delta + \mathcal{O}(\delta).
\end{eqnarray}
\end{widetext}
Note that at this step $r_c$ dropped out just like it did in the previous subsection of this paper, getting replaced by $R$ via the definition \rf{eq:deltadef}. $R$ is defined even with for potentials which extend
all the way to infinity, so the results obtained here, just as elsewhere in this paper, are valid as long as $R$ can be defined via Eq.~\rf{eq:Rdef}  and is finite. 

Repeating the same steps that lead to \rf{eq:partFinals} and \rf{eq:energyFinals} we finally obtain the expression for the number of trapped bosons and the energy of the polaron:

\begin{widetext}
\begin{eqnarray} \label{eq:parama}
N & = &4 \pi n_0 \xi^3 \left[  \delta^\frac{1}{3} - \frac{5}{3 \sqrt{2}} \delta^\frac{2}{3} + 2 \delta \ln \delta  + \dots  + \frac{\xi\delta^\frac{1}{3}}{3a}\left(\delta^\frac{1}{3}-\sqrt{2}\delta^\frac{2}{3} + 4\delta \ln \delta + \dots \right)\right], \cr
E & = & -\frac{\pi n_0\xi}{m}\left[3\delta^\frac{1}{3}-2\sqrt{2}\delta^\frac{2}{3}+4\delta \ln \delta + \dots  + \frac{\xi\delta^\frac{1}{3}}{ a}\left( 2\delta^\frac{1}{3}-\frac{4\sqrt{2}}{3}\delta^\frac{2}{3} +4\delta \ln \delta + \dots \right)\right].
\end{eqnarray}
\end{widetext}

%%%%%%%%%%%%%%%%%%%%%%%%%%%%%%%%%%%%%%%%%%%%%%%%%%%%%%%%%%%%%%%
\begin{figure}
\centering
\includegraphics[width=\columnwidth]{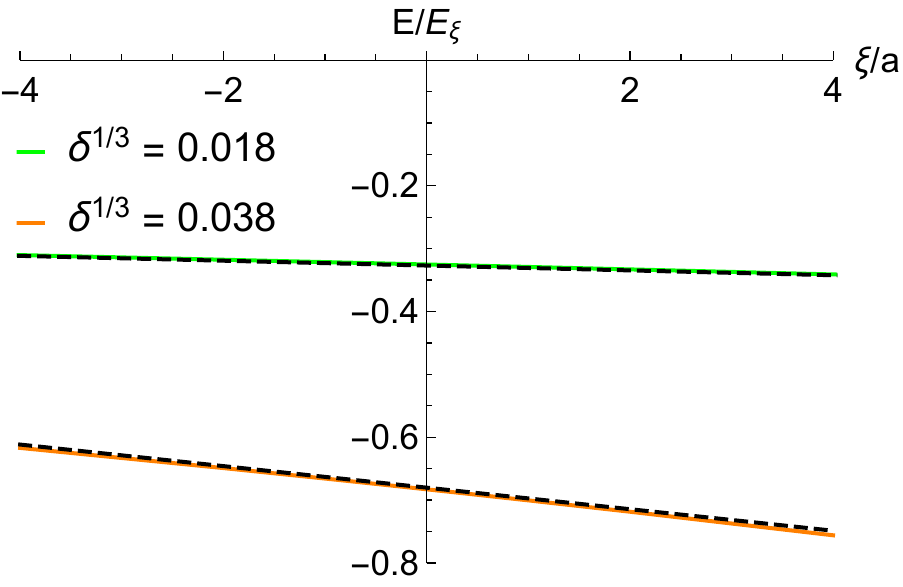}
\caption{\label{fig:oneoverae}
Polaron energy $E$ away from unitarity for the square well impurity-bath potential computed at two different values of $\delta$. The dashed black lines correspond to the analytical expression given by Eq.~\rf{eq:parama}.  Plot as a function of $\xi/a$ in units of $E_\xi = \xi n_0 /(2m)$.} 
\end{figure}
%%%%%%%%%%%%%%%%%%%%%%%%%%%%%%%%%%%%%%%%%%%%%%%%%%%%%%%%%%%%%%%

%%%%%%%%%%%%%%%%%%%%%%%%%%%%%%%%%%%%%%%%%%%%%%%%%%%%%%%%%%%%%%%
\begin{figure}
\centering
\includegraphics[width=\columnwidth]{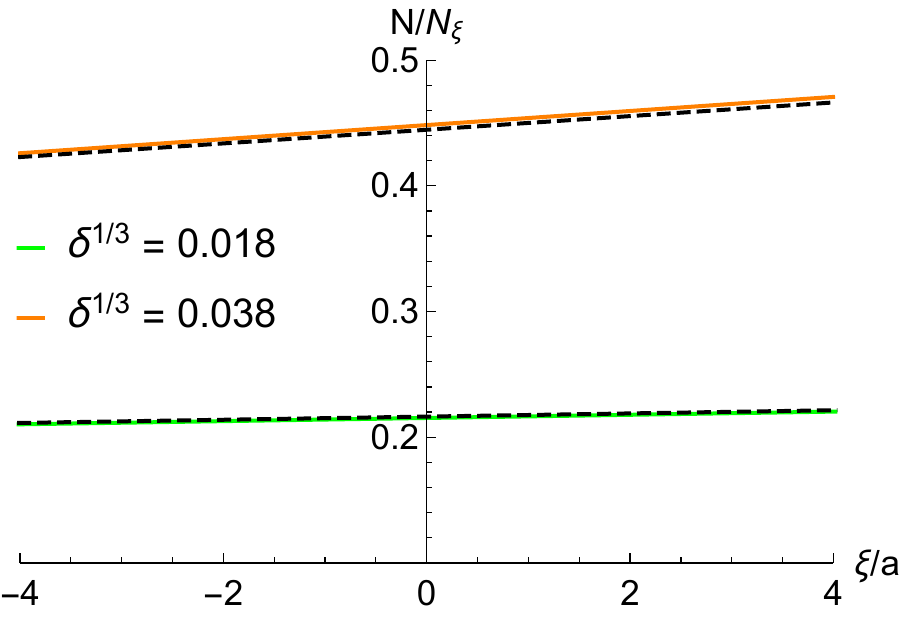}
\caption{\label{fig:oneoverap}
Number of trapped bosons $N$ away from unitarity for the square well impurity-bath potential computed at two different values of $\delta$. The dashed black lines correspond to the analytical expression as given by Eq.~\rf{eq:parama}.  Plot as a function of $\xi/a$ in units of $N_\xi = \xi^3 n_0 $.} 
\end{figure}
%%%%%%%%%%%%%%%%%%%%%%%%%%%%%%%%%%%%%%%%%%%%%%%%%%%%%%%%%%%%%%%

We verify the above expressions by numerically solving the Gross-Pitaevskii equation \rf{eq:GP}, where we choose $U$ to be the square well potential. The analytic expression for the scattering length in this potential is well known, so it is easy to tune the strength of the potential to get the desired value of $a$.  Far away from impurity one can use the asymptotic solution $\phi = 1+\frac{u}{z}$, where $u$ is given by Eq.~\rf{eq:la}. With two values of parameter $A$, $A_\text{min}$ and $A_\text{max}$, our algorithm performs the Newton bisection untill  we find the value of $A$ such that the derivative of the solution at the origin is zero.  The graphs of the polaron energy and the number of trapped bosons  slighly away from unitarity are presented in  Fig.~\ref{fig:oneoverae} and Fig.~\ref{fig:oneoverap}.

\subsubsection{$a/\xi$ corrections for weak potentials}
\label{sec:smallacor}

Now that we have established the machinery for solving the Gross-Pitaevskii equation in both regions, let us revisit the case of the weak potential and find the next order correction to the energy  \rf{eq:energyweak} in powers of the scattering length $a$. Note that the contribution to the energy from the first interval goes in powers of $a/r_c$, while in the second interval in powers of $a/\xi = a\epsilon/{r_c}$, so in principle, the energy expansion should be in powers of $(\frac{a}{r_c})^i \epsilon^j$.  Moreover, from the equation  one can show that the contribution of the region $z \in [0, \epsilon]$ to the energy goes as $\sim \frac{n_0a}{m}(\epsilon + \frac{a}{\xi}\epsilon)$, while the region $z \in [\epsilon, \infty]$ contributes $\sim \frac{n_0a}{m}(1+\frac{a}{\xi})$, so in the limit $\epsilon \rightarrow 0$, only the second region is important. The $\sim a^2$ contribution to the energy comes from $\frac{a^2}{r_c^2}\epsilon^2$ term, so in order to find the energy up to order $a^2$, one needs to find the expression for the amplitude $A$ in the second region up-to order $\epsilon^2$. The structure of  equations in \rf{eq:weakmatch} tells us that the matching of amplitudes determines $A$, while the matching of derivatives determines $\alpha$. Since we need to know only $A$ up to order $\epsilon^2$, we need to correct the first equation by the terms of order $\epsilon^2$, while the second only by the terms of order $\epsilon$. Notice that iteration of the RHS of Eq. \rf{eq:GPrad2} will produce terms of higher order than we need, so this tells us that the result should be independent on the further details of the potential. Having in mind that $A \sim \epsilon$, in the second interval we expand Eq. \rf{eq:itm} up to $\epsilon^2$ terms to produce:

\be \label{eq:weaksyssm}  \begin{matrix} A + \frac{3\ln 3}{2\sqrt{2}}A^2-\sqrt{2}\epsilon A & = & \epsilon(\alpha-1) , \cr
-\sqrt{2} A  & = &  \frac{\alpha}{1-\frac{a}{r_c}} -1. \end{matrix}
\ee

Looking for the solution of this system in the form $\alpha = \alpha_0 + \epsilon \alpha_1$ and $A = \epsilon A_0+\epsilon^2 A_1$, we get

\begin{eqnarray} \label{eq:weakparam}
\alpha & = & 1-\frac{a}{r_c} + \sqrt{2}\frac{a}{r_c}\left(1-\frac{a}{r_c} \right) + \mathcal{O} ( \epsilon^3 \ln \epsilon), \cr
 A  &=&  -\frac{a}{r_c}\epsilon - \frac{(4+3\ln 3)}{2\sqrt{2}}\left(\frac{a}{r_c}\epsilon \right)^2 + \mathcal{O}(\epsilon^3 \ln \epsilon).
\end{eqnarray}

The energy of the polaron is given by

\be
\label{eq:weakenergyeq}
E = -\frac{\lambda}{2} \int d^3 x \left( |\psi|^4-n_0^2 \right) = -2\pi \lambda n_0^2 \xi^3 \int_0^\infty dz z^2 \left(\phi^4-1 \right).
\ee

The region from 0 to $\epsilon$ does not contribute, while the second interval produces:

\be
\label{eq:weakenergyint}
\int_\epsilon^\infty dz z^2 \left( \left(1+\frac{u}{z} \right)^4-1 \right).
\ee

Once again, we use Eq. \rf{eq:itm} with $A$ given by Eq. \rf{eq:weakparam} to compute the integral in Eq. \rf{eq:weakenergyint} and expand the result up-to $\epsilon^2$ terms. Plugging this  
into Eq. \rf{eq:weakenergyeq}, we finally produce

\be \label{eq:enexp2}
E = \frac{2\pi n_0 a}{m}\left(1+\frac{\sqrt{2}a}{\xi} + \dots \right) .
\ee

The second term in the brackets was obtained before in Ref.~\cite{Christensen2015}. This shows that the Gross-Pitaevskii approach discussed above is well suited for studying both weak potentials and potentials tuned to unitarity.

\subsubsection{Quasiparticle properties of the Bose polaron}
Having established the expressions for the solution of the Gross-Pitaevskii equation, now we can compute other key quasiparticle properties such as quasiparticle residue $Z$ and Tan's contact $C$.  
This was already reported in Ref.~\cite{Yegovtsev2021}. Let us  reproduce it here for completeness.

The residue $Z$ quantifies the overlap between the solutions in presence and absence of the impurity. Within the GP treatment, this is given by \cite{Guenther2020} $$ \ln Z= - \int d^3x\, |\psi(x)-\sqrt{n_0}|^2.$$  At unitarity, the above analysis shows that to leading order

$$ \ln Z=-\sqrt{2} \pi n_0\xi^3 \delta^{2/3} + \dots. $$

Another key quasiparticle property is the impurity-bath Tan's contact, which quantifies the change in the polaron energy in response to a small change of the inverse scattering length, $$ C = -8 \pi m \,\frac{\partial E}{\partial \left( a^{-1} \right)}.$$ Using the expression \rf{eq:parama} for energy of the polaron slightly away from unitarity, we get:

\be \label{eq:contact} C = 16 \pi^2 n_0 \xi^2   \left( \delta^{2/3} -\frac{2\sqrt{2}}{3}\delta + 2\delta^{4/3}\ln \delta + \dots \right). \ee
We note that an alternative definition of the contact is based on the impurity-bath density-density correlator evaluated at the core radius, $\tilde C =  16 \pi^2 r_c^2 \left| \psi(r_c) \right|^2$. Using the expression \rf{eq:param} for the amplitude $\beta$ of the Gross-Pitaevskii equation at the core radius, it is easy to show that both expressions for contact agree in the universal regime, where the effects of impurity-boson potentials are captured by a single parameter $\delta$.

\subsection{Summary: the solution of the Gross-Pitaevskii equation}We collect here the main features of the solutions worked out above. 
In the previous subsection we obtained the solution to the Gross-Pitaevskii equation \rf{eq:GP} both when the scattering length in the potential $U$ was small, and when the potential $U$ was tuned to unitarity. In the latter case we could obtain the analytic solution when the range of the potential $R$ was much smaller than the scattering length $\xi$, as an expansion in powers of $R/\xi$
which corresponded to the asymptotic behavior of the polaron in a very weakly interacting condensate.

The method discussed above allows in principle to find the solution for any negative scattering length, from small $a$ to infinite $a$. However, the general expression is cumbersome. Here we only give the answer in the limiting cases of small $a$ and large $a$.

Let us summarize how the solution looks. In all cases, the normalized solution $\phi = \psi/\sqrt{n_0}$ to Gross-Pitaevskii equation
$$ -\frac{\Delta \phi}{2m} + U \phi = \mu \phi \left( 1 - \left| \phi \right|^2 \right)
$$
 is constructed out of a reference function $v(y)$ which is the solution of the zero energy Schr\"odinger equation
\be \label{eq:zeroenergySch} - \frac{d^2 v}{d y^2} + 2 m r_c^2 U(y) \, v =0.
\ee
Here  $y=r/r_c$, $r_c$ is  the radius of the potential, so that $U(r)=0$ for all $r>r_c$ or $y>1$.  $v(0)=0$, while $v(1)=1$. 

When $a$ is small so that \be \label{eq:condw} |a|^3 \ll \xi^2 r_c, \ee  the solution is
\be \label{eq:fullsolw}  \psi(r)  \approx \left\{  \begin{matrix} \frac{r_c}{r}  \left( 1 - \frac a {r_c} \right)  v\left(\frac r {r_c} \right), & r < r_c , \cr 1 - \frac a r  e^{- \frac{\sqrt{2} r}{\xi}},  & r>r_c .   \end{matrix}
\right. \ee
This can be extracted from Eqs.~\rf{eq:BP}, \rf{eq:la} and \rf{eq:solw}. Note that Eq.~\rf{eq:fullsolw} is valid even if $a/r_c > 1$, as long as Eq.~\rf{eq:condw} is valid. 

On the other hand, if 
\be \label{eq:conds} |a|^3 \gg \xi^2 r_c, \ee then the solution is
\be \label{eq:fullsols}  \psi(r)  \approx \left\{  \begin{matrix}  \frac{\xi^{2/3} R^{1/3} }{r}  \, v\left(\frac r {r_c} \right), & r < r_c , \cr 1+\frac{\xi^{2/3} R^{1/3} }{r}   e^{- \frac{\sqrt{2} r}{\xi}},  & r>r_c .   \end{matrix}
\right. \ee
This in turn can be extracted from Eqs.~\rf{eq:within}, \rf{eq:uzerom} and \rf{eq:param}.

Clearly the solution behaves as if when $\left| a \right|$ is increased past $\xi^{2/3} R^{1/3}$, $a$ needs to be simply replaced by $-\xi^{2/3} R^{1/3}$ in the solution.

The energy and the particle number corresponding to this solution have been worked out above in Eqs.~\rf{eq:partFinals} and \rf{eq:energyFinals}, and summarized in Eqs.~\rf{eq:energyas} and
\rf{eq:particleas2}. We also worked out perturbative corrections to the solution for small $a$ and for large $a$ in Sections \ref{sec:smallacor} and \ref{sec:largeacor} respectively,
with the results for the energy in particular summarized in Eqs.~\rf{eq:enexp2} and \rf{eq:parama}. 

%%%%%%%%%

%%%%%%%%%%

%%%%%%%%%%
\section{Fluctuational corrections to the Gross-Pitaevskii equation}
\label{sec:fluct} 
Let us look at the condensate density at and nearby the point where impurity is located. This can be extracted from the solution of the Gross-Pitaevskii equation found above. The density grows as one approaches the center of the polaron. At $r<r_c$, the solution is given by Eq.~\rf{eq:within}. Here $v$ is the solution of the Schr\"odinger equation \rf{eq:sch25}, so it takes values of the order of $1$ for
all $0<r<r_c$. The magnitude of the solution is controlled by the coefficient in front of $v(y)$, which is  of the order of $1/\epsilon^{2/3}$. Therefore, the density of the condensate at the origin
is roughly
$$ n \sim \frac{n_0}{\epsilon^{4/3}}.
$$
We can now estimate the gas parameter at the position of the condensate
$$ n a_B^3 = \frac{n_0 \xi^{4/3} a_B^3}{R^{4/3}} \sim n^{1/3} \frac{a_B^{7/3}}{R^{4/3}}.
$$
The condition that this parameter is small, or
\be \label{eq:newco}  \frac{n_0}{\epsilon^{4/3}} a_B^3 \ll 1,
\ee
 is equivalent, upon expressing $\epsilon$ in terms of $n_0$, $a_B$, and $R$, to
\be \label{eq:critc} R  \gg a_B \left( n_0 a_B^3 \right)^{1/4}.
\ee
Under this condition, the gas remains weakly interacting everywhere, including at the position of the impurity. We expect that the fluctuational corrections remain small under these conditions and
the results obtained from Gross-Pitaevskii equation remain valid. 

Let us briefly note that the criterion \rf{eq:critc} puts a lower bound on $R$. But at the same time, our solution of the Gross-Pitaevskii equation relied on $R$ being much smaller than $\xi$. This criterion can be rewritten as
$$
\delta = \frac{R}{\xi} \sim \frac{R}{a_B} \sqrt{n_0 a_B^3} \ll 1.
$$
Combining this with Eq.~\rf{eq:critc} we find the window of $R$ where our approach works, given by
$$  \left( n_0 a_B^3 \right)^{1/4} \ll \frac{R}{a_B} \ll \frac{1}{\sqrt{n_0 a_B^3}},
$$
as was advertised earlier in Eq.~\rf{eq:conditions}. 

Let us now present the formalism where we formally derive this criterion. We follow the standard approach to fluctuations in a weakly interacting Bose gas. Denote the solution to the Gross-Pitaevskii
equation \rf{eq:GP} as $\psi_0$ and write
\be \psi = \psi_0+ \varphi. \ee
We now substitute this into the action \rf{eq:acc} and expand the action up to the quadratic terms in $\varphi$. These steps are standard in the theory of weakly interacting Bose gas. The only new
aspect of this problem here is the impurity potential $U$ present in our theory. The result of the expansion reads
\begin{eqnarray} S_q &=& \int d^3 x \, d\tau \left[ \bar \varphi \left( \partial_\tau - \mu - \frac{\Delta}{2m}  +U+ 2 \lambda \psi_0^2 \right) \varphi + \right. \cr  &&  \left. \frac \lambda 2 \psi_0^2 \left( \varphi^2 + \bar \varphi^2 \right) 
\right]
\end{eqnarray}
Here $S_q$ denotes the part of the action quadratic in the field $\varphi$. 

This action can now be used to calculate the density of particles which are expelled from the condensate by interactions and are not contained within the solution to the Gross-Pitaevskii equation.
This density must be much smaller than the density contained within the solution to the Gross-Pitaevskii equation in order for the mean field approximation which led to the Gross-Pitaevskii equation to remain valid. 
 In the absence of the potential $U$ this calculation is standard and the answer is given in textbooks, leading to the criterion $n_0 a_B^3 \ll 1$ as the condition for the applicability of the Gross-Pitaevskii
 equation. Our aim is to repeat this calculation in the presence of the impurity potential $U$. 

The density of particles can be accessed via calculating the Green's functions of the field $\varphi$. As is standard in this approach, we define the matrix of Green's functions ${\cal G}$ which include
both normal and anomalous Green's functions, according to
$$ {\cal G}(x_1,x_2,\tau) = - \left( \begin{matrix}  \VEV{\varphi(x_1, \tau) \bar \varphi(x_2, 0)} & \VEV{\varphi(x_1, \tau) \varphi(x_2,0)} \cr \VEV{\bar \varphi(x_1,\tau) \bar \varphi(x_2,0)} & \VEV{\bar \varphi(x_1,\tau) \varphi(x_2,0)} \end{matrix} \right).
$$
It is convenient to work with ${\cal G}$ in the frequency domain, by Fourier transforming it over $\tau$ resulting in ${\cal G}(x_1, x_2, \omega)$. Note that it is not as straightforward to Fourier transform ${\cal G}$ in space, because
the presence of the impurity makes ${\cal G}$ dependent on both $x_1$ and $x_2$ and not just on their difference as would have been the case in the Bose-Einstein condensate without 
the external potential $U$. 

{\cal G} satisfies the following matrix equation
\be {\cal X}(x_1) {\cal G}(x_1,x_2, \omega) = - \delta(x_1- x_2),
\ee where ${\cal X}$ is the  operator-valued matrix
\be\label{eq:grr} {\cal X} = \ee
$$  \left( \begin{matrix}  -i \omega- \frac{\Delta}{2m} - \mu + U + 2 \lambda \psi_0^2 & \lambda  \psi_0^2  \cr
 \lambda  \psi_0^2 &
i \omega - \frac{\Delta}{2m} - \mu + U + 2 \lambda \psi_0^2    \end{matrix} \right). 
$$
In the definition of ${\cal X}$ the Laplacian $\Delta$ implies differentiation over $x_1$, while $\psi_0$ and $U$ depend on $x_1$. 
These equations are difficult to solve exactly  given this explicit $x_1$ dependence. 

Instead of trying to solve these exactly, we employ local density approximation. That includes working with a Wigner transform of ${\cal G}$ defined as
$$ {\cal G} (x, p, \omega) = \int d^3 y \, {\cal G} (x+y/2, x-y/2, \omega) \, e^{-i {\bf y} \cdot {\bf p}}.
$$
This object can be calculated approximately by replacing $\Delta$ Eq.~\rf{eq:grr} by $-p^2$ and working with it as if $V$ and $\psi_0$ are constants, despite their explicit dependence on $x_1$. 
Within this approximation ${\cal G}(x,p, \omega)$ is given simply by the inverse of this matrix derived from ${\cal X}$ 
$$  \left( \begin{matrix}  -i \omega+ \frac{p^2}{2m} - \mu + U + 2 \lambda \psi_0^2 & \lambda  \psi_0^2  \cr
 \lambda  \psi_0^2 &
i \omega + \frac{p^2}{2m} - \mu + U + 2 \lambda \psi_0^2   \end{matrix} \right). 
$$
In particular, the upper left corner of that inverse defines the usual normal Green's function and gives
\be \label{eq:fluc} G(x, p, \omega) =  - \frac{i \omega + \frac{p^2}{2m} - \mu + U(x) + 2 \lambda\psi_0^2(x)}
{\omega^2 + \left( \frac{p^2}{2m} + U(x) + \lambda \psi_0^2(x) \right)^2 - \lambda^2 \psi_0^4}.
\ee

Eq,~\rf{eq:fluc} is by no means exact. It is the result of the local density approximation which relies on $\psi_0(x)$ and $U(x)$ being slow varying on the scale of the coherence length $\xi$, which itself defines the values of the
typical momenta $p$. Because of this, it is sometimes also referred to as gradient expansion, with Eq.~\rf{eq:fluc} being the first term in it. An explicit procedure allowing to compute further terms in this expansion is available, but will not be needed here. 

Now in practice $\psi_0(x)$ varies roughly on the scale of $\xi$, so it is at the limit of the applicability of the local density approximation. $U(x)$ varies on the scale of $r_0 \ll \xi$ so it definitely breaks the
conditions of applicability of Eq.~\rf{eq:fluc}. Nonetheless in the absence of better and as easily accessible technique, and mindful that our goal is not to calculate the fluctuational particle density   but just to estimate it, we will  continue the calculation using Eq.~\rf{eq:fluc}. 

To calculate the density of particles due to fluctuations we need to calculate $\delta n =\VEV{\bar \varphi(x, 0) \varphi (x,0) }$. This can be found by the following succession of steps. First we evaluate
$$ - \lim_{\tau \rightarrow 0^-} \int \frac{d\omega}{2\pi} \, G(x, p, \omega) \, e^{-i \omega \tau},
$$
with the result
$$  \delta n(x,p) = \frac{ \frac{p^2}{2m} - \mu + U(x) + 2 \lambda \psi_0^2(x) - E(p,x)}{2 E(p,x)},
$$ where 
\be \label{eq:fspe} E(p,x) = \sqrt{  \left( \frac{p^2}{2m} - \mu + U(x) + 2 \lambda \psi_0^2(x) \right)^2 - \lambda^2 \psi_0^4}
\ee
can be interpreted as a local in space dispersion of the Bogoliubov quasiparticles. 
The density of particles due to fluctuations can be evaluated using
\be \label{eq:nnum} \delta n (x) = \int \frac{d^3 p}{(2 \pi)^3} \delta n(x,p).
\ee
To evaluate the remaining integral over the momentum we need the explicit form of  $\psi_0$. Let us do it at $U$ tuned to unitarity. We can then use Eq.~\rf{eq:fullsols} for $\psi_0$. 
In particular, with $r$ representing the distance of the point $x$ to the origin, for $r \gg r_c$ we use that
\be \label{eq:fuu}  \lambda \psi_0^2 \approx \mu \left( 1 + \frac{2 \xi^{2/3} R^{1/3}}{r} e^{- \frac{ \sqrt{2} r}{\xi}} + \dots \right)
\ee
At the same time, at these values of $r$ beyond the range of the potential, $U=0$. This can be substituted into the integral in Eq.~\rf{eq:nnum}, and the integral can be  evaluated perturbatively in powers of the second term in Eq.~\rf{eq:fuu} which is small when $r \gg r_c$. Denoting this term as 
$$ \zeta = 2 \frac{\xi^{1/3} R^{2/3}}r e^{- \frac{\sqrt{2} r}{\xi}}
$$
we find
$$ \delta n(x) = \delta n_0 + \frac{(m \mu)^{3/2}}{32} \zeta \ln \zeta.
$$
Here $\delta n_0(x)$ is the standard answer for the density of particles expelled from the condensate in the absence of any external potential, given by $\delta n_0 = - 4 (m \mu)^{3/2}/(3 \pi^2)$. 

Let us compare the density of particles expelled from the condensate to the excess density of particles in the Gross-Pitaevskii solution compared to the density in the absence of impurity. The latter is extracted from Eq.~\rf{eq:fuu} and is simply $n_0 \zeta$. Thus the condition that the density of particles expelled from the condensate is much smaller than the density in the solution to Gross-Pitaevskii solution is
\be \label{eq:pc} (m \mu)^{3/2} \zeta \ln \zeta \ll n_0 \zeta.
\ee
Recalling that $\mu m \sim n_0 a_B$ and neglecting the logarithm in \rf{eq:pc} this is equivalent to
$$ n_0 a_B^3 \ll 1.
$$
Thus we arrive at the conclusion that at $r \gg r_c$ the fluctuational corrections to the solution of the Gross-Pitaevskii equation are small simply if the interactions in the Bose gas are weak. 

Now suppose we bring $r$ close to $r_c$. At this point $U$ is still zero, but the estimate for $\lambda \psi_0$ again taken from Eq.~\rf{eq:fullsols} produces
$$ \lambda \psi_0^2 \sim \frac{\mu}{ \epsilon^{4/3}}.
$$
Evaluating Eq.~\rf{eq:nnum} again produces 
$$ \delta n(x) \sim \frac{( m \mu )^{3/2}}{\epsilon^2}.
$$
The condition that this is much smaller than the density from the Gross-Pitaevskii solution gives
$$ \frac{( m \mu )^{3/2}}{\epsilon^2} \ll \frac{n_0}{\epsilon^{4/3}}.
$$
Again using $m \mu \sim n_0 a_B$, we rewrite this as
$$ \frac{n a_B^3}{\epsilon^{4/3}} \ll 1,
$$ equivalent to Eq.~\rf{eq:newco}. Note that this is stronger than just requiring the gas to be weakly interacting. Thus we reproduced the condition \rf{eq:critc} introduced earlier by 
qualitative reasoning.

Finally, we note that if $r<r_c$, the energy spectrum \rf{eq:fspe} becomes poorly defined for small enough $p \lesssim 1/r_0$ due to $U \sim -1/(m r_0^2)$. This is to be expected as local density approximation should break down for such small $r$.  One way to get around it is to restrict the momenta in the integral in Eq.~\rf{eq:nnum} to be larger than $1/r_0$. However, this would constitute an uncontrolled approximation and would in any case be beyond what we need to obtain the rough estimate of the fluctuations, the task in which we already succeeded by limiting $r$ to $r > r_c$.

 %%%%%%%%%%%%%%%%%%%%%%%%%%%%%%%% 
  %%%%%%%%% Alternative method %%%%%%%%%%%%%%
  %%%%%%%%%%%%%%%%%%%%%%%%%%%%%%%%%

  \section{Finite range boson-boson potential}
  \label{sec:four}

 Here we would like to discuss the effects of a finite-ranged $V_{bb}$ on the solution to the Gross-Pitaevskii equation
  when the impurity potential $U$ is close to the unitary limit. We will see that, as promised earlier, these effects are mild and mostly quantitative, without changing the main qualitative aspects of the
  solution.
  
  A finite ranged $V_{bb}$ changes the Gross-Pitaevskii equation \rf{eq:GP} to read
  \be \label{eq:GPnl} - \frac{\Delta \psi(x)}{2m} + U(x) \psi(x) = \psi(x) \left( \mu  - \int d^3 y \ V_{bb}(x-y) \left| \psi(y) \right|^2 \right).
  \ee 

Remember that in the local theory we described above the expansion was constructed by matching the asymptotic solution far away from the impurity with the solution in the region $r<r_c$. For the theory at hand the asymptotic behaviour of the solution in the regime $r \gg r_c$ is expected to be roughly the same as in the local case, provided that $\epsilon \ll 1$ and $r_b \lesssim r_c$. The leading order solution in the region $r<r_c$ is independent of $r_b$, and since the structure of the higher order perturbative terms depends on the structure of the solutions discussed above, we expect that the effect of nonlocal interaction is to modify the perturbative terms without changing the scaling in $\epsilon$. More formally, the structure of the system of Eqs. \rf{eq:syssm} is such that all numeric coefficients are of the order of unity and this allows to solve the system in powers of $\epsilon^{1/3}$. The nonlocal generalization would introduce more terms that should depend on a $r_b/r_c$. Provided that $r_b/r_c \sim 1$, we can still solve the resultant system of equation in powers of $\epsilon^{1/3}$, so the expansion for energy and the number of the trapped bosons would have a similar form as in the local scenario, but the values of the coefficient cannot be determined analytically.  As such, we performed extensive numerical evaluations of this equation. 

First, the energy of the polaron is given by
\be  \label{eq:energynl}  E = -\frac{1}{2}\int d^3x \ d^3y \ V_{bb}(x-y)(|\psi(x)|^2|\psi(y)|^2-n_0^2). \ee

Here $n_0$ is the uniform density of the gas in the absence of the impurity. The number of trapped bosons is given by the same expression as in the local Gross-Pitaevskii theory

\be \label{eq:npnl} N = \int d^3x \left[|\psi|^2-n_0 \right]. \ee

In order to make contact with our previous discussion of the zero-ranged boson-boson potentials, it is is convenient to define $V_{bb}$ in such a way that we formally retrieve expression given by \rf{eq:vbb} once the range $r_b$ of the boson-boson interaction is taken to zero. This also implies that $n_0 = \frac{\mu}{\lambda}$ as for the local case. In our numerical study we will use a purely repulsive Gaussian boson-boson potential of the form
\be \label{eq:Vbbdef} V_{bb} = \frac{\lambda}{\pi^{3/2}r_b^3}e^{-\frac{(x-y)^2}{r_b^2}}. \ee

An advantage of using this potential is that for spherically symmetric solutions one can explicitly perform the angular integration on the RHS of Eq.~\rf{eq:GPnl}, so the resultant equation becomes one-dimensional. For the potential given in \rf{eq:Vbbdef} this gives

$$\int d^3 y V_{bb}(x-y) \left| \psi(y) \right|^2 \rightarrow $$ 
\be \label{eq:symme} \frac{\lambda }{\sqrt{\pi} r_b x}\int_0^\infty y dy \, e^{-\frac{(x-y)^2}{r_b^2}}\left(1-e^{-\frac{4xy}{r_b^2}} \right) |\psi(y)|^2. \ee 

Since the Gross-Pitaevskii theory treats boson-boson potential in the Born-approximation, the relation between the coupling constant $\lambda$ and the boson-boson scattering length $a_B$ is the same as in the local theory: $\lambda = 4\pi a_B/m$.

To find the ground state of  Eq.~\rf{eq:GPnl}, we perform evolution in imaginary time $\tau$:

$$ -\partial_\tau \psi = - \frac{1}{2m} \left( \frac{\partial^2 \psi}{\partial x^2}  + \frac 2 x \frac{\partial \psi}{\partial x} \right)  + (U(x)- \mu)  \psi(x) + $$
\be \label{eq:GPnltau}    \frac{\lambda  \psi(x)}{\sqrt{\pi} r_b x}\int_0^\infty y dy \, e^{-\frac{(x-y)^2}{r_b^2}}\left(1-e^{-\frac{4xy}{r_b^2}} \right) |\psi(y)|^2.  \ee

 Here the initial state at $\tau=0$ is taken to be the solution of the Gross-Pitaevskii equation \rf{eq:GP} given by Ref.~\rf{eq:fullsols}. As $\tau \rightarrow \infty$, this produces the solution to the Eq.~\rf{eq:GPnl}.  We discretize the spatial part of the above equation, so that \rf{eq:GPnltau} becomes a system of coupled nonlinear equations. The continuum limit is retrieved by extrapolating numerical results to zero step-size.

\begin{figure}
\centering
\includegraphics[width=\columnwidth]{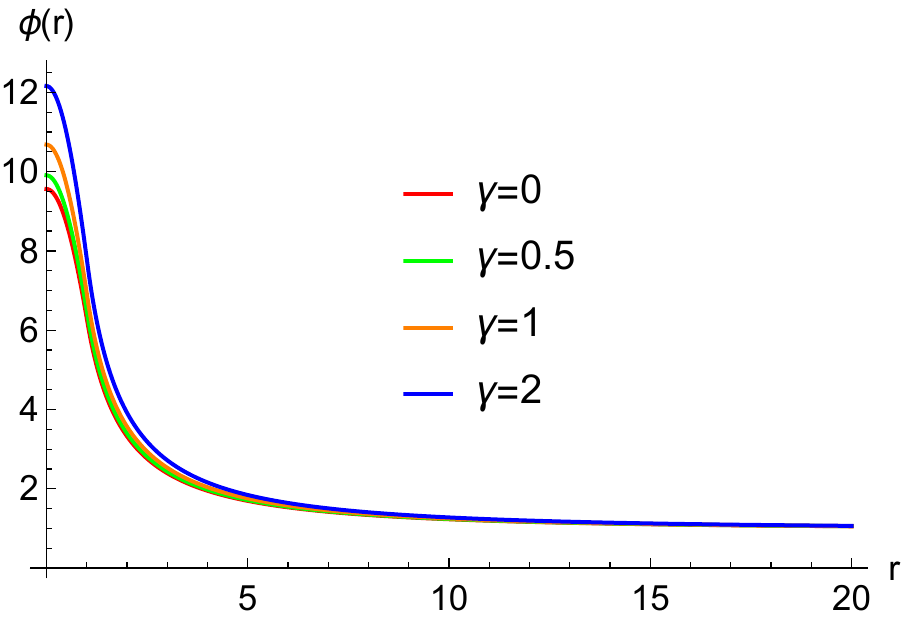}
\caption{\label{fig:nlGPe}
Solution of the nonlocal GP equation for various values of $\gamma$ at fixed $\epsilon=0.05$.
} 
\end{figure}

It is convenient to introduce new dimensionless parameter $\gamma = \frac{r_b}{r_c}$, so that in the limit $\gamma \rightarrow 0$, we retrieve the original Gross-Pitaevskii equation. Carrying
out the numerical procedure, we find that for the fixed value of $\epsilon$, as we increase $\gamma$ from zero to some finite value such that $\gamma \sim 1$, the solutions to the Eq.~\rf{eq:GPnl} evolve in a way that the amplitude of the solution at the origin is an increasing function of $\gamma$ and all solutions approach the same asymptotic value far away from the impurity as indicated in  Fig.~\ref{fig:nlGPe}. Eq. \rf{eq:npnl} tells us that for a fixed $r_c$ and $\xi$, increasing $r_b$ increases the number of trapped particles which in turn lowers the energy of the polaron.

%%%%%%%%%%%%%%%%%%%%%%%%%%%%%%%%%%%%%%%%%%%%%%%%%%%%%%%%%%%%%%%
%%%%%%%%%%%%%%%%%%%%%%%%%%%%%%%%%%%%%%%%%%%%%%%%%%%%%%%%%%%%%%%

\begin{figure}
\centering
\includegraphics[width=\columnwidth]{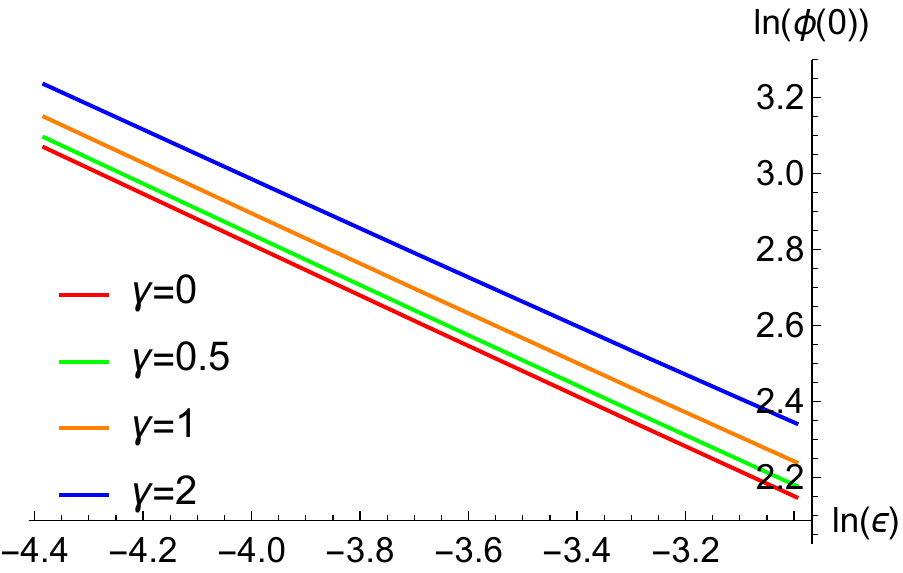}
\caption{\label{fig:nlGPeAmp}
Scaling of the amplitude of the nonlocal GP equation at the origin for various values of $\gamma$. All lines are parallel to the $\gamma=0$ line which has $\sim \epsilon^{-2/3}$ scaling predicted by local theory, indicating that this scaling survives when $\gamma$ is small. 
}
\end{figure}

We test this assertion numerically by studying the scaling of the amplitude at the origin and show that it indeed scales as $\sim \epsilon^{-2/3}$ as is shown in  Fig.~\ref{fig:nlGPeAmp}.

%%%%%%%%%%%%%%%%%%%%%%%%%%%%%%%%%%%%%%%%%%%%%%%%%%%%%%%%%%%%%%%
%%%%%%%%%%%%%%%%%%%%%%%%%%%%%%%%%%%%%%%%%%%%%%%%%%%%%%%%%%%%%%%

%%%%%%%%%%%%%%%%%%%%%%%%%%%%%%%%%%%%%%%%%%%%%%%%%%%%%%%%%%%%%%%
\begin{figure}
\centering
\includegraphics[width=\columnwidth]{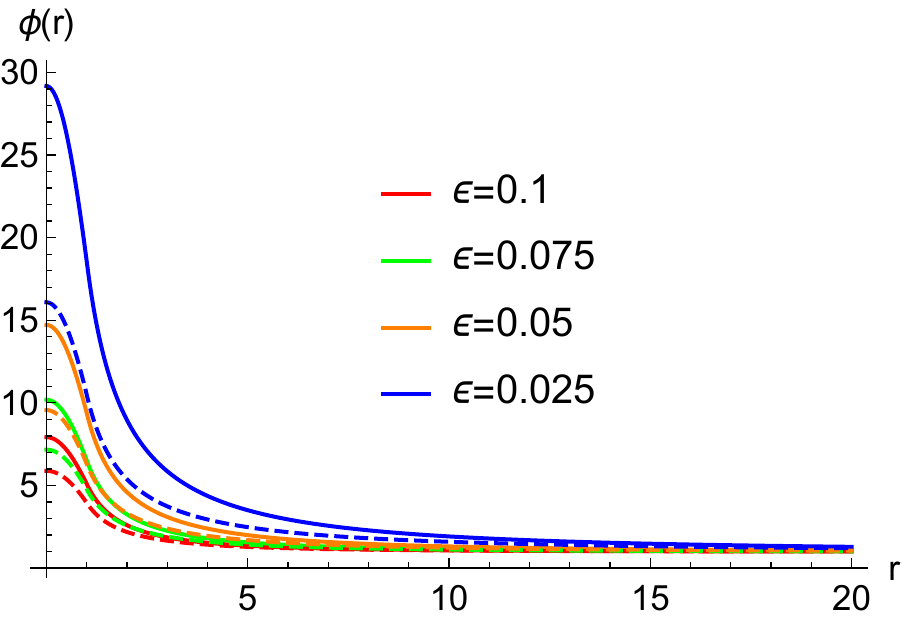}
\caption{\label{fig:nlGPe0range}
Amplitudes of the wavefunction close to the impurity. Solid lines show the result of the non-local theory with boson-boson range $r_b = r_c /(4\epsilon)$, while dashed lines are obtained setting $r_b=0$ (i.e., are obtained from the usual local GP equation).
}
\end{figure}
%%%%%%%%%%%%%%%%%%%%%%%%%%%%%%%%%%%%%%%%%%%%%%%%%%%%%%%%%%%%%%%

%%%%%%%%%%%%%%%%%%%%%%%%%%%%%%%%%%%%%%%%%%%%%%%%%%%%%%%%%%%%%%%
\begin{figure}
\centering
\includegraphics[width=\columnwidth]{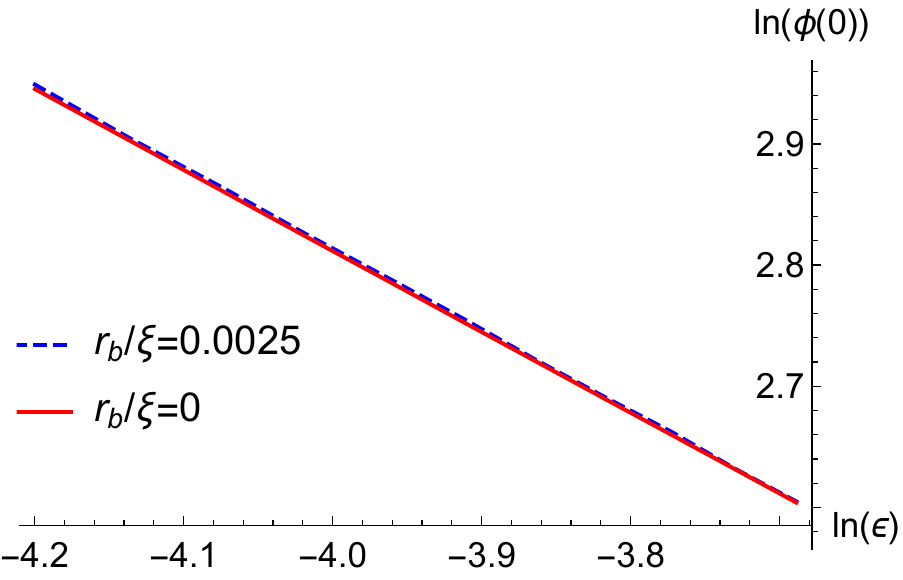}
\caption{\label{fig:nlGPerboverxi}
Amplitude of the wavefunction obtained in the case $ r_b \ll r_c$. The dashed line shows the result for $\gamma = 0.0025/\epsilon$, while the solid line shows the result of the local GP equation ($\gamma$=0).
}
\end{figure}
%%%%%%%%%%%%%%%%%%%%%%%%%%%%%%%%%%%%%%%%%%%%%%%%%%%%%%%%%%%%%%%

The discussion above shows that introducing a finite-ranged boson-boson interaction introduces quantitative but non qualitative changes to the problem. It is interesting now to study what happens if one tries to shrink the other range in the problem, the one between the bosons and the impurity. To do so, we decrease $r_c$ at fixed $r_b$ and $\xi$. In the language of $\epsilon$ and $\gamma$, this corresponds to the limit $\epsilon \rightarrow 0$, $\gamma \rightarrow \infty$ with $\epsilon\gamma$ kept fixed, which is the limit of a contact boson-impurity interaction. Fig.~\ref{fig:nlGPe0range} shows that making the range of the impurity potential smaller, while keeping the range of the boson interaction and the healing length fixed, significantly increases the density at the origin. We expect that the density at the origin will diverge in the limit of contact impurity interactions, however since we cannot access regime of very small value of $\epsilon$ and large values of $\gamma$ numerically, we leave the case of the contact impurity interaction for future work.

In the intermediate regime, where $r_b \ll r_c$ and $r_b/\xi$ fixed, we studied the scaling of the amplitude at the origin and found that it follows the $\sim \epsilon^{-2/3}$ scaling as is shown in Fig. ~\ref{fig:nlGPerboverxi}. Thus in the limit  $r_b \ll r_c$ and even $r_b \sim r_c$ we expect that solutions to the Gross-Pitaevskii equation would qualitatively look similar to the $r_b=0$ case explored analytically. Numerics shows that for $\gamma =1$ energy of the polaron is about 10 percent lower than the energy of $\gamma=0$ case.

 %\section{Comparison with the available numerical data}

  \section{Expanding about the unperturbed condensate}
  \label{sec:flat}
  An alternative method of evaluating the functional integral \rf{eq:free} exists. Instead of calculating it by the saddle point approximation, that is, by minimizing the action and expanding
  it about the minimum, we expand about
  a flat unperturbed condensate 
  \be \psi = \sqrt{n_0} = \sqrt{\frac{\mu}{\lambda}},
  \ee
  ignoring at first the polaron potential $U$. 
Unlike the approach discussed in the previous section, this method is an uncontrolled approximation. It turns out that it produces correct answer only if the scattering length $a$ of the potential $U$
describing impurity-boson interactions is sufficiently small. However, in principle this method produces an answer (given below in Eq.~\rf{eq:energyflat}) which one could attempt to use
for arbitrary values of $a$.
In fact, this  expression for the polaron energy already appeared in the literature before, obtained using a variety of different techniques. Here we argue that this answer breaks down as $a$ becomes larger, and is entirely incorrect as the potential is tuned to unitarity where $a$ is infinity, despite some claims in the literature to the contrary. 

To apply this method to our problem, it is convenient to begin by introducing a Hubbard-Statonovich field. To do that consistently, we will focus only on the case when the attractive potential $U$ is negative everywhere, $U(r)<0$. We can then write
\be e^{\int d\tau d^3 x \, U \bar \psi \psi } = \int {\cal D} d \, e^{\int d\tau d^3 x \, \left( \bar d \psi + \bar \psi d + U^{-1} \bar d d \right)}.
\ee 
With this done, we apply the expansion about the unperturbed condensate into the functional integral defined in Eq.~\rf{eq:free}. We write
\be \psi = \sqrt{n_0} + \varphi,
\ee Then we expand the action in powers of $\varphi$, keeping only quadratic terms, which corresponds to the Bogoliubov approximation to the weakly
interacting Bose gas.   We emphasize that it is this step which is approximate, and as we will discuss later, this step breaks down if the potential is strongly attractive. 
 
 This produces the following functional integral 
  \begin{widetext} 
\be \int {\cal D}  \varphi \, {\cal D} d \, \exp\left( - \int d^3x d\tau \left[ \bar \varphi \left( \pp{\tau} - \mu - \frac{\nabla}{2m} \right) \varphi + \frac{\mu}{2} \left( \varphi^2 + \bar \varphi^2 \right) + \mu \bar \varphi \varphi 
- \sqrt{n_0} (d+ \bar d) - \bar d \varphi - \bar \varphi d - U^{-1} \bar d d
 \right] \right).
\end{equation} 
Crucially, this integral is Gaussian, so can be calculated exactly. This is what we are going to do now. 
We first integrate out  bosonic fluctuations $\varphi$. 
With the standard definition of the Fourier transform
\be d(\tau, {\bf r}) =T  \sum_{{\bf p}, \omega} d(\omega, {\bf p}) \, e^{-i \omega \tau + i {\bf p}\cdot { \bf r}},
\ee
we work in frequency-momentum domain and find the effective action
\be \label{eq:accc}  \frac  S V = - \sqrt{\rho} \left( d(0) + \bar d(0) \right) + T \sum_{\omega, \bf p} G_{\rm n}  \bar d(\omega, {\bf p}) d(\omega,{\bf p}) + \frac T 2 \sum_{\omega, \bf p} G_{\rm a}  \left( d (\omega,{\bf p}) d(-\omega, -{\bf p})+\bar d(\omega, {\bf p}) \bar d(-\omega, -{\bf p}) \right)+
\ee $$
T \sum_{{\bf p} {\bf q}} U^{-1}({\bf p} - {\bf q} ) \bar d(\omega,{\bf p}) d(\omega, {\bf q} ).
$$
\end{widetext}
Here $\omega = 2\pi n T$ are the usual bosonic Matsubara frequencies, $d(0)$ above implies $d$ calculated at both zero frequency and momentum, and
 $G_{\rm n}$ and $G_{a}$ are normal and anomalous Green's functions of the Bose gas in the Bogoliubov approximation,
\be G_{\rm n} = - \frac{ i \omega + \frac{p^2}{2m} + \mu}{\omega^2 + \left( \frac{p^2}{2m} + \mu \right)^2 - \mu^2},
\ee
\be G_{a} = \frac{\mu}{\omega^2 + \left( \frac{p^2}{2m} + \mu \right)^2 - \mu^2}.
\ee
We note that at nonzero frequencies $S$  is quadratic in fields, while at zero frequency it includes a linear term. Therefore, the most important contribution will come from the zero frequency terms.
To integrate over zero frequency terms, we minimize the action, solve the resulting equation and substitute back into the action. Minimizing gives the following equation
\be \label{eq:s1} - \frac{\sqrt{\rho}}{T} \, U({\bf p}) + d({\bf p})+ \ee
$$ \sum_{\bf q} U({\bf p}-{\bf q})  \left[  G_{\rm n}({\bf q}) d({\bf q}) + G_{\rm a}({\bf q}) \bar d(-{\bf q})  \right] =0,
$$ where everything is now at zero frequency.
Look for a solution  in terms of a real function $d({\bf r})$, or in other words $d({\bf p}) = \bar d(-{\bf p})$ and denote
\be f({\bf p})  = T \frac{d({\bf p})}{2 \mu + \frac {p^2}{2m}}.
\ee
Then the equation \rf{eq:s1} can be rewritten in the position space as
\be \label{eq:s3}  {\sqrt{n_0}}  \, U + \left( -\frac{\Delta}{2m} + 2\mu + U \right) f =0.
\ee
We now note that this is nothing but the expansion of the Gross-Pitaevskii equation \rf{eq:GP} about the flat solution. In other words, Eq.~\rf{eq:s3} can be obtained from Eq.~\rf{eq:GP} by substituting
$\psi \approx \sqrt{n_0} + f$ and expanding in powers of $f$. This should already tell us that the applicability of the method we are using is limited, since $f$ is not always small. 

Ignoring this issue for now, we note that once Eq.~\rf{eq:s3} is solved, the solution needs to be substituted back into Eq,~\rf{eq:accc} to find the action
\be \label{eq:Sans} S = - V \sqrt{n_0} \, d(0),
\ee
which constitutes the answer obtained in this calculation.

We now know enough about the solution to the full Gross-Pitaevskii equation to easily solve Eq.~\rf{eq:s3}. For $0<r<r_c$ we can neglect $\mu$. The solution is then simply
$$ \frac{f}{\sqrt{n_0}} = - 1 + B \frac{v}{r},
$$
where $v$, as defined earlier, is the solution of the zero energy Schr\"odinger equation \rf{eq:zeroenergySch}, normalized as explained after Eq.~\rf{eq:zeroenergySch}, and $B$ is yet some unknown constant. 

At $r>r_c$ the potential vanishes and the solution to Eq.~\rf{eq:s3} reads
$$ \frac{f}{\sqrt{n_0}}  = \frac A r e^{- \sqrt{2} r/\xi}.
$$
We now match these by taking advantage of the Bethe-Peierls boundary conditions satisfied by $v$
$$ - r_c +B = A, \ - A \sqrt{2}/\xi = -1 + B/(r_c-a).
$$
Solving these in the limit $r_c \ll \xi$, we find
$$ A = \frac{\xi}{\sqrt{2} -\xi/a}, \ B = \frac{\xi}{\sqrt{2} - \xi/a}.
$$
We now calculate the action according to Eq.~\rf{eq:Sans},
$$ S = - \frac{2 \mu  \sqrt{n_0}}{T} \int d^3 x \, f(x).
$$
The leading contribution to the integral produces
$$ S =  \frac{2 \pi a n_0  V}{T m \left( 1-a\sqrt{2}/\xi \right)}.
$$
From here we finally deduce that, since $S = EV/T$, 
\be \label{eq:energyflat} E = \frac{2 \pi a n_0  }{ m \left( 1-\sqrt{2}\, a /\xi \right)}.
\ee
This is a very appealing expression, and not surprisingly it appeared previously in the literature (see Ref.~\cite{Shchadilova2016}, in particular their Eq.~(4) as well as their Supp. Mat.). The first two terms of its expansion in powers of $a/\xi$ are given by
\be \label{eq:energyweakexp} E \approx \frac {2 \pi a n_0}m \left( 1+ \frac{\sqrt{2} \, a}{\xi} \right) + \dots.
\ee 
The first of these is the standard well-known weak potential answer which we  obtained before in Eq.~\rf{eq:energyweak}.
The second term is the correction to that in the expansion in powers of $a/\xi$ which we already obtained earlier in Eq.~\rf{eq:enexp2} and which goes back to the work in Ref.~\cite{Christensen2015}. 
Both of these are definitely correct. 

It is very tempting to use the expression for the energy \rf{eq:energyflat} at larger $|a|$ as well, all the way to $a$ going to infinity where it produces $E = -\sqrt{2} \pi n_0 \xi/m$.
However, there is no reason to expect that this would be correct. See Appendix below for a toy problem which illustrates why only the 
first two terms of Eq.~\rf{eq:energyflat} as given in Eq.~\rf{eq:energyweakexp} can be trusted (see also Ref.~\cite{Parish2021} for further discussions of Eq.~\rf{eq:energyflat}).

%%%%%%%%%%%%%%%%%%%%%%%%%%%%%%%%%%%%%%%%%%%%%%%
%%%%%%%%%%%%%%%%%%%%%%%%%%%%%%%%%%%%%%%%%%%%%%%
%%%%%%%%%%%%%%%%%%%%%%%%%%%%%%%%%%%%%%%%%%%%%%%
%%%%%%%%%%%%%%%%%%%%%%%%%%%%%%%%%%%%%%%%%%%%%%%
%%%%%%%%%%%%%%%%%%%%%%%%%%%%%%%%%%%%%%%%%%%%%%%
%%%%%%%%%%%%%%%%%%%%%%%%%%%%%%%%%%%%%%%%%%%%%%%

\section{Impurity-boson interactions supporting a bound state}
\label{sec:seven}

 Let us now briefly explore the regime where the potential $U$ is so strongly attractive that it now supports a bound state.  To do this,  we mostly follow the results reported  in our earlier publication \cite{Yegovtsev2021}. 

As the potential $U$ increases in strengths beyond unitary limit, the scattering length $a$ in such a potential is now positive.  This implies that it now has a bound state with binding energy 
$$ E_{\rm binding} = -\nu = -\frac 1 {2 m a^2},
$$ $a$ becomes positive. If $a$ becomes sufficiently small so that
the relationship \rf{eq:weaks} holds again, simple arguments give the energy and the number of trapped particles of the polaron as
\be \label{eq:bound}  E  \sim -\frac{m R^3 \nu^2}{a_B}, \ N \sim \frac{m R^3 \nu}{a_B}.
\ee 
where the precise coefficients now depend on the details of the potential $U$. (A subtlety in trying to use Eq.~\rf{eq:trapped} for Eq.~\rf{eq:bound}  is that an additional term in $E$ suppressed by a power of a  factor of $\delta$ compared to
what is presented in Eq.~\rf{eq:bound} is needed to recover the expression for $N$.)

Indeed, suppose $N$ bosons get trapped in this bound state, then its energy is 
$$ E = - N \nu + g \frac{N^2} 2,
$$ where the self-repulsion constant $g$ can be estimated as \be g \sim \frac{\lambda }{R^3}.\ee Minimizing $E$ with respect to $N$ we find Eqs.~\rf{eq:bound}. This solution can also be obtained from the Gross-Pitaevskii equation if one notes that 
it corresponds to the density of bosons being $n \sim N/R^3 \sim \nu/\lambda$, and that results in the nonlinear term in the Gross-Pitaevskii equation $\lambda \left| \psi \right|^2 \psi \sim \nu \psi$, thus turning Gross-Pitaevskii equation approximately into the Schr\"odinger equation at energy $-\nu$, whose solution will roughly follow the bound state solution of the Schr\"odinger equation. Such solution of the Gross-Pitaevskii equation, which would fix the coefficients in Eq.~\rf{eq:bound}, can only be found numerically and the answer, which will fix the numerical coefficients in Eq.~\rf{eq:bound}, will be highly dependent on the details of the potential $U$. 

It is straightforward to estimate $$ n a_B^3 \sim  (a_B/a)^2 \ll 1,$$ justifying the use of the Gross-Pitaevskii 
equation as long as $a \gg a_B$.

\section{Conclusions}
\label{sec:conclusions}

We presented here a theory of impurities in weakly interacting Bose condensates, attractively interacting with bosons which formed the condensate. Our theory is based on Gross-Pitaevskii equation in external potential.
We demonstrate that the approximation of Gross-Pitaevskii equation is valid as long as the range of the potential is not too small and not too large, as described by the criteria \rf{eq:conditions}. The theory remains valid for arbitrary  impurity-boson scattering length, including in the unitary limit where the scattering length goes to infinity. 

We demonstrate that for weakly interacting Bose gases it is possible to solve the corresponding Gross-Pitaevskii equation \rf{eq:GP} analytically. Therefore, the weakness of intra-boson interactions 
play dual role in our theory: they allow for the analytic solution of Gross-Pitaevskii equation, and they suppress quantum fluctuations about the Gross-Pitaevskii solution, if the properly defined range of the potential is finite and lies within the interval \rf{eq:conditions}. 
Within this theory we found the binding energy of the polaron and the number of bosons that become trapped in the vicinity of the impurity. In the regime of impurity-boson interactions at unitarity, they are given by Eqs.~\rf{eq:energyas} and \rf{eq:particleas2}. Perturbing away from unitarity slightly we find Eq.~\rf{eq:parama}. For generic attractive potential with negative scattering length which is neither small nor large the answer can be found analytically via the formalism developed here, but we found its analytic expression too cumbersome to present here. 

We work with intra-bosons interactions of zero range. We explore the effects of finite range of intra-boson interactions and demonstrate that taking it into account does not appreciably change our results. 

The questions which we have not yet addressed include the behavior of the polaron at finite momentum, including the determination of its effective mass. We leave these questions for future work. 

\vspace{5mm}
\begin{acknowledgments}
We acknowledge inspiring and insightful discussion with G.~E.~Astrakharchik, J. Levinsen, and M. Parish.
PM was supported by grant PID2020-113565GB-C21 funded by MCIN/AEI/10.13039/501100011033, and EU FEDER Quantumcat.
 This work was also supported by the Simons Collaboration on Ultra-Quantum Matter, which is a grant from the Simons Foundation (651440, VG, NY), and by the National Science Foundation under Grant No. NSF PHY-1748958.
\end{acknowledgments}

\appendix*
\section{Failure of the expansion about the unperturbed condensate: a toy problem}
\label{sec:appendix}
To illustrate the method used in this paper, we study the following toy problem. First we would like to evaluate the integral 
\be \label{eq:sad}  I= \int_0^\infty  dx \, e^{a x^2 - bx^4},
\ee
in case where $b$ is small and $a$ is positive. We will evaluate it by the saddle point method. 
The saddle point is
\be x_0 = \sqrt{\frac{a}{2b}}.
\ee
Writing $x=x_0 + y$, and $a x^2 - b x^4 = a^2/(4 b) - 2 a (x-x_0)$, we find
\be \label{eq:apint} I \approx \sqrt{\frac{\pi}{2a}}  \,  {\exp\left( {{\frac{a^2}{4b}}} \right)} .
\ee
This is the standard answer to $I$, produced as an exponential of the power series in $b$. 

Now  suppose we would like to evaluate the following integral ($a>0$, $c>0$, $b$ is small). 
\be f =\frac{  \int_0^\infty dx \, e^{a x^2 - b x^4} e^{c x^2}} { \int_0^\infty dx  \, e^{a x^2 - b x^4} }.
\ee
On the one hand, from Eqs~\rf{eq:sad} and \rf{eq:apint}, the answer is
\be \label{eq:correct} f \approx  e^{\frac{2 a c + c^2 }{4b}} \sqrt \frac {a}{a+c}.
\ee
On the other hand, we attempt to evaluate $f$ by taking the following steps, mimicking the approach employed in Section~\ref{sec:flat}. We introduce a Hubbard-Stratonovich variable $y$
\be f = \frac{1}  { \int_0^\infty dx  \, e^{a x^2 - b x^4} }  { \int_0^\infty dx  \, \int_{-\infty}^\infty \frac{dy}{\sqrt{4 \pi c} }  \, e^{a x^2 - b x^4 + x y - \frac{y^2}{4 c}}}.
\ee
To evaluate the integral over $x$ in the numerator of the expression above, we use
\be x =  \sqrt{\frac{a}{2b}} + z,
\ee 
and expand in powers of $z$. Note that this is not a legitimate way to approach this problem, however, this is what was done in Section~\rf{sec:flat} when the functional integral
was expanded about n unperturbed condensate. 
We find
\be  f \approx \frac{1}  { \int_0^\infty dx  \, e^{a x^2 - b x^4} }  { \int_{-\infty}^\infty    \frac{dz dy}{\sqrt{4 \pi c} }  \, e^{\frac{a^2}{4b} - 2 a z^2 + \left( \sqrt{\frac a {2b}} + z \right) y - \frac{y^2}{4 c}}}
=
\ee $$  \sqrt{\frac{ a}{2\pi^2 c }}{ \int_{-\infty}^\infty dz   d y  \, e^{ - 2 a z^2 + \left( \sqrt{\frac a {2b}} + z \right) y - \frac{y^2}{4 c}}}.
$$
Evaluating the integral over $z$ gives
\be f \approx \frac{1}{\sqrt{4 \pi c}}  \int_{-\infty}^\infty dy \, e^{ \sqrt{\frac{a}{2b}} y + \frac{y^2}{8a} - \frac{y^2}{4c}} = \sqrt{\frac{ 2 a}{2a - c}} e^{\frac{a^2  c}{b(2a - c)}}.
\ee
Compare this with the correct answer \rf{eq:correct}. Expanding the leading asymptotic of the answer in the exponential in powers of $c$ if $c$ is small,
\be \frac{a^2  c}{b(2a - c)} \approx  \frac{a c }{2b } + \frac{c^2}{4b} + \frac{c^3}{8 a b} + \dots, 
\ee
so the first two terms do indeed coincide with the  correct expression in the exponential
\be \frac{2 a c + c^2}{4b}.
\ee
However, the rest of the terms have nothing to do with the correct answer. 

This is an indication that in Eq.~\rf{eq:energyflat} only the first two terms in the expansion in powers of $a$, as given in Eq.~\rf{eq:energyweakexp}, could be trusted.

\bibliography{UnitaryPolaron}

\end{document}